\documentclass[12pt,a4paper]{article}
\usepackage{amssymb}
\usepackage{amsfonts}
\usepackage[onehalfspacing]{setspace}
\usepackage{multirow}
\usepackage{pstricks}

\newtheorem{theorema}{Theorem}

\newtheorem{theorem}{Theorem}[section]
\newtheorem{corollary}[theorem]{Corollary}

\newtheorem{lemma}[theorem]{Lemma}
\newtheorem{proposition}[theorem]{Proposition}

\newenvironment{proof}[1][Proof]{\noindent\textbf{#1.} }{\ \rule{0.5em}{0.5em}}
\newtheorem{example}[theorem]{Example}
\let\oldexample\example
\renewcommand{\example}{\oldexample\normalfont}
\newtheorem{remark}[theorem]{Remark}
\let\oldremark\remark
\renewcommand{\remark}{\oldremark\normalfont}
\input{tcilatex}

\begin{document}

\title{Selling Multiple Correlated Goods: Revenue Maximization and Menu-Size
Complexity\thanks{%
Previous versions, titled \textquotedblleft The Menu-Size Complexity of
Auctions": August 2012; April 2013 (The Hebrew University of Jerusalem,
Center for Rationality DP-637 and http://arxiv.org/abs/1304.6116); November
2017. Presented at the 2013 ACM\ Conference on Electronic Commerce. Research
partially supported by a European Research Council Advanced Investigator
grant (Hart), by an Israel Science Foundation grant (Nisan), and by a Google
grant (Nisan). We thank Motty Perry and Phil Reny for introducing us to the
subject and for many insightful discussions, and the referees and editors of
this journal for their many useful comments. A\ presentation that covers
some of this work is available at \texttt{%
http://www.ma.huji.ac.il/hart/abs/2good-p.html}}}
\author{Sergiu Hart\thanks{%
The Hebrew University of Jerusalem (Federmann Center for the Study of
Rationality, Department of Economics, and Institute of Mathematics).\quad 
\emph{E-mail}: \texttt{hart@huji.ac.il} \quad \emph{Web site}: \texttt{%
http://www.ma.huji.ac.il/hart}} \and Noam Nisan\thanks{%
The Hebrew University of Jerusalem (Federmann Center for the Study of
Rationality, and School of Computer Science and Engineering), and Microsoft
Research. \emph{E-mail}: \texttt{noam@cs.huji.ac.il} \quad \emph{Web site}: 
\texttt{http://www.cs.huji.ac.il/\symbol{126}noam}}}
\maketitle

\begin{abstract}
We consider the well known, and notoriously difficult, problem of a single
revenue-maximizing seller selling two or more heterogeneous goods to a
single buyer whose private values for the goods are drawn from a (possibly
correlated) known distribution, and whose valuation is additive over the
goods. We show that when there are two (or more) goods, \emph{simple
mechanisms}---such as selling the goods separately or as a bundle---\emph{%
may yield only a negligible fraction of the optimal revenue. }This resolves
the open problem of Briest, Chawla, Kleinberg, and Weinberg (\emph{JET}
2015) who prove the result for at least three goods in the related setup of
a unit-demand buyer. We also introduce the \emph{menu size} as a simple
measure of the complexity of mechanisms, and show that the revenue may
increase polynomially with menu size and that no bounded menu size can
ensure any positive fraction of the optimal revenue. The menu size also
turns out to \textquotedblleft pin down\textquotedblright\ the revenue
properties of deterministic mechanisms.
\end{abstract}

\tableofcontents

%TCIMACRO{%
%\TeXButton{References Without Numbers}{\def\@biblabel#1{#1\hfill}
%\def\thebibliography#1{\section*{References}
%\addcontentsline{toc}{section}{References}
%\list
%{}{
%\labelwidth 0pt
%\leftmargin 1.8em
%\itemindent -1.8em
%\usecounter{enumi}}
%\def\newblock{\hskip .11em plus .33em minus .07em}
%\sloppy\clubpenalty4000\widowpenalty4000
%\sfcode`\.=1000\relax\def\baselinestretch{1}\large \normalsize}
%\let\endthebibliography=\endlist}}%
%BeginExpansion
\def\@biblabel#1{#1\hfill}
\def\thebibliography#1{\section*{References}
\addcontentsline{toc}{section}{References}
\list
{}{
\labelwidth 0pt
\leftmargin 1.8em
\itemindent -1.8em
\usecounter{enumi}}
\def\newblock{\hskip .11em plus .33em minus .07em}
\sloppy\clubpenalty4000\widowpenalty4000
\sfcode`\.=1000\relax\def\baselinestretch{1}\large \normalsize}
\let\endthebibliography=\endlist%
%EndExpansion

\section{Introduction\label{s:intro}}

Are complex auctions better than simple ones? Myerson's (1981) classic
result (see also Riley and Samuelson 1981 and Riley and Zeckhauser 1983)
shows that if one is aiming to maximize revenue when selling a single good,
then the answer is \textquotedblleft no.\textquotedblright\ The optimal
auction is very simple, allocating the good to the highest bidder (using
either first or second price) as long as he bids above a single
deterministically chosen reserve price.

However, when selling multiple goods the situation turns out to be much more
complex. There has been significant work both in economics and in computer
science\footnote{%
See Section \ref{sus:literature} for a literature survey.} showing that, for
selling multiple goods, simple auctions are no longer optimal. Specifically,
it is known that randomized auctions may yield more revenue than
deterministic ones, and that bundling the goods may yield higher (or lower)
revenue than selling each of the goods separately. This is true even in the
very simple setting where there is a single buyer.

In this paper we consider such a simple setting: a single seller, who aims
to maximize his expected revenue, sells two or more heterogeneous goods to a
single buyer whose private values for the goods are drawn from an arbitrary
(possibly correlated) but known prior distribution, and whose value for
bundles is additive over the goods in the bundle. Since we are considering
only a single seller, this work may alternatively be interpreted as dealing
with the monopolistic pricing of multiple goods.\footnote{%
See Appendix \ref{ap:n-buyers} for the extension of our results from the
single-buyer to the multiple-buyer setting.}

In our previous paper, Hart and Nisan (2017, originally circulated in 2012),
we considered the setup where the buyer's values for the different goods are 
\emph{independent}, in which case we showed that simple mechanisms are \emph{%
approximately} optimal: selling each good separately (deterministically) for
its optimal price extracts a constant fraction of the optimal revenue. In
this paper we show that the picture changes completely when the valuations
of the goods are \emph{correlated}, in which case \textquotedblleft
complex\textquotedblright\ mechanisms can become arbitrarily better than
\textquotedblleft simple\textquotedblright\ ones.

The setup is that of $k$ goods, whose valuation to the single buyer is given
by a random variable $X=(X_{1},X_{2},...,X_{k})$ with values in $\mathbb{R}%
_{+}^{k};$ we emphasize that we allow for arbitrary dependence between the
coordinates of $X.$ The buyer's valuation for a bundle of goods is additive
over the goods; thus, for example, getting the first two goods is worth $%
X_{1}+X_{2}$ to the buyer$.$ We denote by \textsc{Rev}$(X)$ the optimal
revenue achievable by any mechanism for selling $k$ goods to an additive
buyer with a random valuation $X.$

Consider first the case of just two goods, i.e., $k=2.$ When the valuations
of the two goods are independent (i.e., $X_{1}$ and $X_{2}$ are independent
random variables), Hart and Nisan (2017) showed that selling the goods
separately---each one at its optimal one-good price---is guaranteed to yield
at least $50\%$ of the optimal revenue, a bound that was later improved to $%
62\%$ by Hart and Reny (2017). This can be stated in terms of the
\textquotedblleft Guaranteed Fraction of Optimal Revenue" (\textsc{GFOR})%
\footnote{%
See Hart and Nisan (2017): given a class of mechanisms $\mathcal{N}$ and a
class of valuations $\mathbb{X},$ we define \textsc{GFOR}$(\mathcal{N};%
\mathbb{X})$ as the maximal fraction of the optimal revenue that can be
achieved by mechanisms in $\mathcal{N}$ for any valuation in $\mathbb{X}$
(cf. Section \ref{sus:gfor-mob}).} as%
\[
\text{\textsc{GFOR}}(\text{\textsc{separate}; }2\text{ \emph{independent
goods}})\geq \frac{\sqrt{e}}{\sqrt{e}+1}\approx 0.62.
\]

How does this fraction change when the two goods need not be independent?
Our first result is that it drops all the way down to zero:%
\[
\text{\textsc{GFOR}}(\text{\textsc{separate}; }2\text{\emph{\ goods}})=0.
\]%
Indeed, we show that\footnote{%
A stronger result is in fact proved; see Section \ref{s:results} for precise
statements of the main results.}

\begin{quote}
\emph{\textbf{For every }}$\varepsilon >0$ \emph{\textbf{there exists a
two-good random valuation }}$X$ \emph{\textbf{with values in }}$[0,1]^{2}$ 
\emph{\textbf{such that}} 
\[
\text{\textsc{SRev}}(X)<\varepsilon \cdot \text{\textsc{Rev}}(X), 
\]
\end{quote}

\noindent where \textsc{SRev} stands for the \textquotedblleft separate
revenue" achievable by selling the goods separately. Thus, for correlated
goods, selling separately may yield only an arbitrarily small fraction of
the optimal revenue. We emphasize that, while we provide specific such
random valuations $X,$ none of the constructions in this paper are
knife-edge or pathological (see Remark \ref{r:pathologic}).

This suggests considering the other one-dimensional mechanism, namely, that
of selling the two goods as a bundle. That does not help: the guaranteed
fraction of optimal revenue is still zero; i.e.,%
\[
\text{\textsc{GFOR}}(\text{\textsc{bundled}; }2\text{\emph{\ goods}})=0. 
\]%
In fact, even the larger class of all \textquotedblleft deterministic"
mechanisms---in which the seller sets a price for each good separately as
well as a price for the bundle---does not fare any better:%
\begin{equation}
\text{\textsc{GFOR}}(\text{\textsc{deterministic}; }2\text{\emph{\ goods}}%
)=0.  \label{eq:gfor(det,2)=0}
\end{equation}

This immediately extends to any number of goods $k\geq 2$ (just add $k-2$
goods with zero valuation):%
\begin{equation}
\text{\textsc{GFOR}}(\text{\textsc{deterministic}; }k\geq 2\text{\emph{\
goods}})=0.  \label{eq:gfor(det,k)=0}
\end{equation}

While these results are new in the case of $k=2$ goods, they have already
been established for $k\geq 3$ goods in the related model of a \emph{%
unit-demand} (instead of additive) buyer---i.e., a buyer who wants to get 
\emph{only one} of the $k$ goods---by Briest, Chawla, Kleinberg, and
Weinberg (2015, originally circulated in 2010); the case of two goods was
left open, with some partial results indicating that \textsc{GFOR} may be
bounded away from zero for $k=2.$ While the unit-demand model and our
additive model are different, they are closely related: the various revenues
in the two models are within constant factors of one another (see Appendix %
\ref{ap:unit demand} for precise statements). On the one hand, this implies
that our result (\ref{eq:gfor(det,k)=0}) for $k\geq 3$ goods follows from
the above-mentioned result of Briest et al. (2015); on the other hand, our
result (\ref{eq:gfor(det,2)=0}) solves their open problem for $k=2$: there
is an infinite gap between the deterministic revenue and the optimal revenue
in the unit-demand model, already for two goods.

What these results say is that allowing for probabilistic outcomes, where
the buyer gets some goods with probabilities that are strictly between $0$
and $1,$ makes a huge difference in terms of revenue. But is it really the
probabilistic vs. deterministic distinction that matters here? A
deterministic mechanism for $k$ goods consists of setting prices for
nonempty subsets of goods and thus provides to the buyer at most $2^{k}-1$
nonzero outcomes to choose from. Suppose we were to limit the seller to
provide the same number, i.e., $2^{k}-1,$ of outcomes, but allow these
outcomes to be probabilistic; would that significantly increase the revenue?
The answer is that it would not! As we will see, the guaranteed fraction of
optimal revenue remains zero for \emph{any} fixed bound on the number of
outcomes.

Formally, we define the \emph{menu size} of a mechanism to be the number of
possible outcomes of the mechanism, where an outcome (or \textquotedblleft
menu entry\textquotedblright ) specifies for each good $i$ the probability $%
q_{i}$ that it is allocated to the buyer, together with the payment $s$ that
the buyer pays to the seller;\footnote{%
See Dobzinski (2011) for an earlier use of menu size in the context of
combinatorial auctions.} it turns out to be convenient not to count the
\textquotedblleft zero\textquotedblright\ outcome of getting nothing and
paying nothing (this outcome is always available, as it corresponds to the
individual rationality or participation constraint). It is easy to see, and
well known, that in our setting any mechanism can be put into the normal
form of offering a fixed menu and letting the buyer choose among these menu
entries. Notice that while deterministic mechanisms for $k$ goods can have a
menu size of at most $2^{k}-1$ (since each $q_{i}$ must be $0$ or $1$),
randomized mechanisms can have an arbitrarily large, even infinite, menu
size. Let \textsc{Rev}$_{[m]}(X)$ denote the optimal revenue achievable by
mechanisms whose menu size is at most $m.$ For a single good, $k=1$, the
characterization of optimal mechanisms of Myerson (1981) implies that 
\textsc{Rev}$_{[1]}(X)$ is already the same as the optimal \textsc{Rev}$(X)$%
, but this is no longer true for more than a single good: the revenue may
strictly increase as we allow the menu size to increase.

Our general result is that for any fixed $m,$ mechanisms that have at most $%
m $ menu entries cannot guarantee any positive fraction of the optimal
revenue:%
\begin{equation}
\text{\textsc{GFOR}}(\text{\textsc{menu size}}\leq m\text{; }k\text{ \emph{%
goods}})=0  \label{eq:gfor(m)=0}
\end{equation}%
for any number of goods $k\geq 2$ and any menu size $m\geq 1.$ Thus, having
a large set of possible outcomes---a large menu from which the buyer
chooses, according to his valuation (or type)---seems to be the crucial
attribute of the high-revenue mechanisms: it enables the sophisticated
screening between different buyer types that is required for high-revenue
extraction. As stated above, taking $m=2^{k}-1$ yields result (\ref%
{eq:gfor(det,k)=0}), which suggests that (\ref{eq:gfor(det,k)=0}) is not
driven by the mechanisms being deterministic, but rather by their being
limited in the number of outcomes that they can offer.

Result (\ref{eq:gfor(m)=0}) says that it does not matter exactly how
\textquotedblleft simple" mechanisms are defined; as long as their menu size
is bounded (which is natural, as unbounded menu size can hardly be
considered simple\footnote{%
See the discussion in Section \ref{sus:menu size} on other complexity
measures that do not use the \textquotedblleft normal form" menu
representation.}), we have

\begin{quote}
\emph{\textbf{For multiple goods, simple mechanisms \textit{cannot}
guarantee any positive fraction of the optimal revenue.}}
\end{quote}

The fact that all simple mechanisms look equally \textquotedblleft bad" when
compared to the optimal revenue-maximizing mechanism does not however
preclude some mechanisms from being better than others in terms of their
revenues. This leads us to compare mechanisms by taking as a benchmark the
simplest \emph{basic} revenue (rather than the optimal revenue), which we
take to be \textsc{Rev}$_{[1]}(X),$ the revenue that is achievable from 
\emph{a single} take-it-or-leave it offer (i.e., a single menu entry); as we
will see in Section \ref{sus:rev-m}, this basic revenue turns out to be
nothing other than the revenue from selling the bundle of all goods at its
optimal price, which we denote by \textsc{BRev}$(X)$.\footnote{%
The \textquotedblleft \textsc{B}" in \textsc{BRev} may thus stand also for
\textquotedblleft \textsc{B}asic."} Thus, given a mechanism $\mu $ we define
the \textquotedblleft Multiple of Basic revenue" of $\mu $, or \textsc{MoB}$%
(\mu )$ for short, to be the maximum, over all (relevant) valuations $X,$ of
the ratio of the revenue that $\mu $ extracts from $X$ to the basic revenue 
\textsc{Rev}$_{[1]}(X)$ from $X$ (the definition is then extended to classes
of mechanisms by taking the maximum over the mechanisms in the class). Thus 
\textsc{MoB}$(\mu )$ measures how many times better the revenue from $\mu $
can be relative to the basic revenue.

The \textsc{MoB} measure turns out to be a useful tool for the analysis:
first, it is given by a simple explicit formula (see Theorem \ref{th:beta});
second, finding a sequence of mechanisms whose \textsc{MoB} goes to infinity
is equivalent to proving that \textsc{GFOR}$($\textsc{bounded}$)=0$ (see
Lemma \ref{l:gfor-mob}(i)); and third, for any class $\mathcal{N}$ of
mechanisms with a bounded \textsc{MoB}---such as deterministic mechanisms,
or mechanisms with bounded menu size---the result that \textsc{GFOR}$(%
\mathcal{N})=0$ follows immediately from \textsc{GFOR}$($\textsc{bounded}$)=0
$\ (see Lemma \ref{l:gfor-mob}(ii)).

We also show that the relation between \textsc{MoB} and menu size is
polynomial (see Theorem \ref{th:rev-m}); that \textsc{MoB} of deterministic
mechanisms is exponential in the number of goods (specifically, for many
goods, i.e., large $k,$ \textsc{MoB} of deterministic mechanisms is
essentially the same as \textsc{MoB} of mechanisms with the same menu size,
i.e., $2^{k}-1$; see Theorem \ref{th:drev}); and, finally, that \textsc{MoB}
of separate-selling mechanisms is linear in the number of goods
(specifically, it equals the number of goods $k$; see Theorem \ref{th:srev}%
). 

Our results thus show that the menu size, although it is just a simple and
crude measure of the complexity of mechanisms, is nevertheless strongly
related to revenue-extraction capabilities.

\subsection{Organization of the Paper\label{sus:paper}}

In Section \ref{sus:literature} immediately below we briefly go over some of
the related literature. Section \ref{s:prelim} presents our model, defines
the menu-size complexity measure and the revenue comparison tools \textsc{%
GFOR} and \textsc{MoB}, and provides some preliminary results. The main
results are then stated in Section \ref{s:results}. Section \ref{s:beta}
deals with the \textsc{MoB} measure, which is then used in Sections \ref%
{s:better} and \ref{s:gap} to construct valuations that prove our results;
see Section \ref{sus:proof-outline} for a detailed guide to the proofs.
Section \ref{s:additive menu} studies separate selling, and introduces a
more refined \textquotedblleft additive menu size\textquotedblright\
complexity measure. We conclude in Section \ref{s:bounded} with positive
approximation results for the case where the valuations are in a bounded
domain. Additional results are relegated to the appendices: the computation
of \textsc{MoB} for two-good deterministic mechanisms (Appendix \ref{ap:beta}%
); the use of the separate-selling revenue, instead of the bundled revenue,
as the \textquotedblleft basic" revenue (Appendix \ref{ap:gamma}); the
relations between our setup and the unit-demand setup (Appendix \ref{ap:unit
demand}); and the multiple-buyer case (Appendix \ref{ap:n-buyers}).

\section{Literature\label{sus:literature}}

We briefly survey some of the existing work on these issues.

The realization that maximizing revenue with multiple goods is a complex
problem has had a long history in economic theory and more recently in the
computer science literature as well. McAfee and McMillan (1988) identify
cases where the optimal mechanism is deterministic. However, Thanassoulis
(2004) and Manelli and Vincent (2006) found a technical error in the paper
and presented counterexamples.\footnote{%
See Hart and Reny (2015) for a simple and transparent such example, together
with a discussion of why this phenomenon can occur only when there is more
than one good.} These papers contain good surveys of the related work within
economic theory, with more recent studies by Fang and Norman (2006), Pycia
(2006), Manelli and Vincent (2007, 2012), Jehiel, Meyer-ter-Vehn, and
Moldovanu (2007), Lev (2011), Pavlov (2011), Hart and Reny (2015). In the
past few years algorithmic work on these types of topics has been carried
out. One line of work shows that for discrete distributions the optimal
mechanism can be found by linear programming in rather general settings:
Briest, Chawla, Kleinberg, and Weinberg (2010/2015),\footnote{%
By \textquotedblleft 2010/2015" we mean \textquotedblleft conference
publication in 2010 and journal publication in 2015."} Cai, Daskalakis, and
Weinberg (2012a), Alaei, Fu, Haghpanah, Hartline, and Malekian (2012).
Another line of work deals with optimal mechanisms for multiple goods in
various settings: Daskalakis, Deckelbaum, and Tzamos (2013, 2017),
Giannakopoulos (2014), Giannakopoulos and Koutsoupias (2014), Menicucci,
Hurkens, and Jeon (2015), Tang and Wang (2017). Yet another line of work
attempts to approximate the optimal revenue by simple mechanisms in various
settings, where simplicity is defined qualitatively: Chawla, Hartline, and
Kleinberg (2007), Chawla, Hartline, Malec, and Sivan (2010), Chawla, Malec,
and Sivan (2010), Alaei, Fu, Haghpanah, Hartline, and Malekian (2012), Cai,
Daskalakis, and Weinberg (2012b). In this line of research, Hart and Nisan
(2012/2017) consider mechanisms that sell the goods either separately or as
a single bundle to be simple mechanisms, and show that when the values of
the goods are \emph{independently} distributed then a nontrivial fraction of
the optimal revenue can be ensured by simple mechanisms. This was followed
by various improved approximation results for independently distributed
goods: Li and Yao (2013), Babaioff, Immorlica, Lucier, and Weinberg (2014),
Yao (2014), Rubinstein and Weinberg (2015), Babaioff, Nisan, and Rubinstein
(2018). By contrast, Briest, Chawla, Kleinberg, and Weinberg (2010/2015)
consider deterministic mechanisms to be simple, and, in the unit-demand
setting with at least $3$ \emph{correlated} goods, prove that deterministic
mechanisms cannot ensure any positive fraction of the revenue of general
mechanisms.

Approaches to quantifying the complexity of mechanisms are studied by
Balcan, Blum, Hartline, and Mansour (2008), Dughmi, Han, and Nisan (2014),
Morgenstern and Roughgarden (2015); we discuss these in Section \ref%
{sus:menu size}. Since the circulation in 2013 of early versions of the
present paper there has been additional work on menu-size complexity; see
Babaioff, Gonczarowski, and Nisan (2017), Gonczarowski (2017), the tutorial
of Goldner and Gonczarowski (2018), and the references there.

When the valuations are bounded, the approximation of auctions and
mechanisms by various discretizations is studied by Hartline and Koltun
(2005), Balcan, Blum, Hartline, and Mansour (2008) (where the construction
is attributed to Nisan), Briest, Chawla, Kleinberg, and Weinberg
(2010/2015), Daskalakis and Weinberg (2012), Dughmi, Han, and Nisan (2014);
see the discussion following the statement of Theorem \ref{th:bounded} in
Section \ref{s:results}.

\section{Preliminaries\label{s:prelim}}

\subsection{The Model\label{sus:model}}

The basic model is standard, and the notation follows our previous paper
Hart and Nisan (2017), which the reader may consult for further details (see
also Hart and Reny 2015).

One seller (or \textquotedblleft monopolist") is selling a number $k\geq 1$
of goods\ (or \textquotedblleft items," \textquotedblleft objects," etc.) to
one buyer.

The goods have no value or cost to the seller. Let $x_{1},x_{2},...,x_{k}%
\geq 0$ be the values of the goods to the buyer. The value of getting a set
of goods is \emph{additive}: getting the subset $I\subseteq \{1,2,...,k\}$
of goods is worth $\sum_{i\in I}x_{i}$ to the buyer (and so, in particular,
the buyer's demand is \emph{not} restricted to one good only). The valuation
of the goods is given by a random variable $X=(X_{1},X_{2},...,X_{k})$ that
takes values in $\mathbb{R}_{+}^{k}$ (we thus assume that valuations are
always nonnegative); we will refer to $X$ as a $k$\emph{-good} \emph{random
valuation.} The realization $x=(x_{1},x_{2},...,x_{k})\in \mathbb{R}_{+}^{k}$
of $X$ is known to the buyer, but not to the seller, who knows only the
distribution $F$ of $X$ (which may be viewed as the seller's belief); we
refer to a buyer with valuation $x$ also as a buyer of \emph{type }$x$. The
buyer and the seller are assumed to be risk neutral and to have quasi-linear
utilities.

The objective is to \emph{maximize} the seller's (expected) \emph{revenue}.

As was well established by the so-called \textquotedblleft Revelation
Principle\textquotedblright\ (starting with Myerson 1981; see for instance
the book of Krishna 2010), we can restrict ourselves to \textquotedblleft
direct mechanisms" and \textquotedblleft truthful
equilibria.\textquotedblright\ A direct\emph{\ mechanism} $\mu $ consists of
a pair of functions\footnote{%
All functions in this paper are assumed to be Borel-measurable (cf. Hart and
Reny 2015, footnotes 10 and 48).} $(q,s),$ where $q=(q_{1},q_{2},...,q_{k}):%
\mathbb{R}_{+}^{k}\rightarrow \lbrack 0,1]^{k}$ and $s:\mathbb{R}%
_{+}^{k}\rightarrow \mathbb{R},$ which prescribe the \emph{allocation} of
goods and the \emph{payment}, respectively. Specifically, if the buyer
reports a valuation vector $x\in \mathbb{R}_{+}^{k},$ then $q_{i}(x)\in
\lbrack 0,1]$ is the probability that the buyer receives good\footnote{%
When the goods are infinitely divisible and the valuations are linear in
quantities, $q_{i}$ may be alternatively viewed as the \emph{quantity} of
good $i$ that the buyer gets.} $i$ (for $i=1,2,...,k$), and $s(x)$ is the
payment that the seller receives from the buyer; we refer to $(q(x),s(x))$
as an \emph{outcome.} When the buyer reports his value $x$ truthfully, his
payoff is\footnote{%
The scalar product of two $n$-dimensional vectors $y=(y_{1},...,y_{n})$ and $%
z=(z_{1},...,z_{n})$ is $y\cdot z=\sum_{i=1}^{n}y_{i}z_{i}$.} $%
b(x)=\sum_{i=1}^{k}q_{i}(x)x_{i}-s(x)=q(x)\cdot x-s(x),$ and the seller's
payoff is $s(x).$

The mechanism $\mu =(q,s)$ satisfies \emph{individual rationality} (\textbf{%
IR)} if $b(x)\geq 0$ for every $x\in \mathbb{R}_{+}^{k};$ it satisfies \emph{%
incentive compatibility} (\textbf{IC}) if $b(x)\geq q(\tilde{x})\cdot x-s(%
\tilde{x})$ for every alternative report $\tilde{x}\in \mathbb{R}_{+}^{k}$
of the buyer when his value is $x,$ for every $x\in \mathbb{R}_{+}^{k}$.

The (expected) revenue of a mechanism $\mu =(q,s)$ from a buyer with random
valuation $X,$ which we denote by $R(\mu ;X),$ is the expectation of the
payment received by the seller; i.e., $R(\mu ;X)=\mathbb{E}\left[ s(X)\right]
.$ We now define

\begin{itemize}
\item \textsc{Rev}$(X),$ the \emph{optimal revenue}, is the maximal revenue
that can be obtained: \textsc{Rev}$(X)=\sup_{\mu }R(\mu ;X),$ where the
supremum is taken over all IC and IR mechanisms $\mu .$
\end{itemize}

\noindent As seen in Hart and Nisan (2017), when maximizing revenue we can
limit ourselves without loss of generality to IR and IC mechanisms that
satisfy in addition the \emph{no positive transfer} (\textbf{NPT}) property: 
$s(x)\geq 0$ for every $x\in \mathbb{R}_{+}^{k}$ (and so $%
s(0,0,...,0)=b(0,0,...,0)=0).$

\textbf{From now on we will assume that all mechanisms }$\mu $\textbf{\ are
given in direct form, i.e., }$\mu =(q,s),$\textbf{\ and that they satisfy
IR, IC, and NPT. }

When there is only one good, i.e., when $k=1,$ Myerson's (1981) result is
that 
\begin{equation}
\text{\textsc{Rev}}(X)=\sup_{p\geq 0}p\cdot \mathbb{P}\left[ X\geq p\right]
=\sup_{p\geq 0}p\cdot \mathbb{P}\left[ X>p\right] =\sup_{p\geq 0}p\cdot
(1-F(p)),  \label{eq:one good}
\end{equation}%
where $F$ is the cumulative distribution function of $X.$ Thus, there are
optimal mechanisms where the seller \textquotedblleft posts" a price $p$ and
the buyer buys the good for the price $p$ whenever his value is at least $p$%
; in other words, the seller makes the buyer a \textquotedblleft
take-it-or-leave-it" offer to buy the good at price $p.$

Besides the maximal revenue \textsc{Rev}$(X)$, we are also interested in
what can be obtained from certain classes of mechanisms.

\begin{itemize}
\item \textsc{SRev}$(X),$ the \emph{separate revenue}, is the maximal
revenue that can be obtained by selling each good separately. Thus 
\[
\text{\textsc{SRev}}(X)=\text{\textsc{Rev}}(X_{1})+\text{\textsc{Rev}}%
(X_{2})+...+\text{\textsc{Rev}}(X_{k}). 
\]

\item \textsc{BRev}$(X),$ the \emph{bundling revenue}, is the maximal
revenue that can be obtained by selling all the goods together in one
\textquotedblleft bundle." Thus 
\[
\text{\textsc{BRev}}(X)=\text{\textsc{Rev}}(X_{1}+X_{2}+...+X_{k}). 
\]

\item \textsc{DRev}$(X),$ the \emph{deterministic revenue},\textsc{\ }is the
maximal revenue that can be obtained by deterministic mechanisms; these are
the mechanisms in which every good $i=1,2,...,k$ is either fully allocated
or not at all: $q_{i}(x)\in \{0,1\}$ for all valuations $x\in \mathbb{R}%
_{+}^{k}$ (rather than $q_{i}(x)\in \lbrack 0,1])$.
\end{itemize}

\noindent While the separate and bundling revenues are obtained by solving
one-dimensional problems (using (\ref{eq:one good})), for each good in the
former, and for the bundle in the latter, the deterministic revenue is a
multidimensional problem.

\subsection{Menu and Menu Size\label{sus:menu size}}

Given a $k$-good mechanism $\mu =(q,s),$ we define its \emph{menu} as the
range of its nonzero outcomes, i.e., 
\[
\text{\textsc{menu}}(\mu ):=\{(q(x),s(x)):x\in \mathbb{R}_{+}^{k}\}%
\backslash \{(0,0,...,0),0)\}\subset \lbrack 0,1]^{k}\times \mathbb{R}_{+}
\]%
(we ignore the zero outcome, $((0,0,...,0),0),$ which is always included
without loss of generality as it corresponds to the IR constraint\footnote{%
We thus slightly depart from Hart and Reny (2015) (where the menu includes
the zero outcome as well); this yields simple relations (such as Proposition %
\ref{p:revm}) between menu size and revenue.}). We will refer to each
outcome in the menu as a \emph{menu entry}. Conversely, any set of outcomes $%
M\subset \lbrack 0,1]^{k}\times \mathbb{R}_{+}$ generates a mechanism $\mu
=(q,s)$ with $(q(x),s(x))\in \arg \max_{(g,t)}(g\cdot x-t)$ where $(g,t)$
ranges over $M\cup \{((0,0,...,0),0)\},$ whose menu is included in\footnote{%
As some outcomes in $M$ may never be chosen; it will be convenient at times
to ignore this and refer to such a $\mu $ as a mechanism with menu $M$.} $M$
(the mechanism is well defined up to tie-breaking; see Hart and Reny 2015
for more details).

The \emph{menu size} of a mechanism $\mu $ is defined as the cardinality of
its menu, i.e., the number of elements of \textsc{menu}$(\mu ),$ which may
well be infinite: 
\[
\text{\textsc{menu size}}(\mu ):=|\text{\textsc{menu}}(\mu )|.
\]%
Since a menu cannot contain two entries $(g,t)$ and $(g,t^{\prime })$ with
the same allocation $g\in \lbrack 0,1]^{k}$ but with different payments $t$
and $t^{\prime }$ (if, say, $t^{\prime }>t$ then $(g,t^{\prime })$ will
never be chosen, as $(g,t)$ is strictly preferred to it by every buyer
type), the menu size is identical to the cardinality of the set of nonzero
allocations; i.e.,%
\[
\text{\textsc{menu size}}(\mu )=|\{q(x):x\in \mathbb{R}_{+}^{k}\text{ and }%
q(x)\neq (0,0,...,0)\}|.
\]%
The corresponding revenue is:

\begin{itemize}
\item \textsc{Rev}$_{[m]}(X),$ the \textquotedblleft \emph{menu-size-}$m$" 
\emph{revenue}, is the maximal revenue that can be obtained by mechanisms
whose menu size is at most $m.$
\end{itemize}

We will refer to the menu-size-$1$ revenue \textsc{Rev}$_{[1]}$ as the \emph{%
basic revenue}: it is the revenue achievable from a single
take-it-or-leave-it offer$.$

Interestingly, Babaioff, Gonczarowski, and Nisan (2017) have recently shown
that the \emph{communication complexity} of a mechanism is precisely the
base $2$ logarithm of its menu size.

Menu size is clearly a very crude measure of the complexity of a mechanism.
In particular, it is based on the \textquotedblleft normal" form of the
mechanism (namely, a menu), and so it ignores the fact that a large menu may
well be representable in a very succinct manner. Such an approach, namely, a 
\emph{Kolmogorov complexity} notion, is used by Dughmi, Han and Nisan
(2014). The \emph{additive menu size}, a refinement of menu size that we
introduce in Section \ref{s:additive menu}, is also a step in this
direction. Another approach is based on \emph{learning-like} notions of
\textquotedblleft dimension": see Balcan, Blum, Hartline, and Mansour (2008)
and Morgenstern and Roughgarden (2015). The advantage of our menu-size
measure is that it is simple, it is defined for each mechanism separately
(rather than for classes of mechanisms), and, as we will see below,\footnote{%
See the tutorial of Goldner and Gonczarowski (2018) and the references there
for additional such results.} it provides useful connections to
revenue-extraction capabilities.

\subsection{Basic Results on Menu Size\label{sus:rev-m}}

We provide here a few simple and immediate relations concerning menu-size
complexity and revenue.

\begin{proposition}
\label{p:revm}For every $k\geq 2$ and every $k$-good random valuation $X$ we
have

(i)%
\begin{equation}
\text{\textsc{Rev}}_{[1]}(X)=\text{\textsc{BRev}}(X);  \label{eq:rev1}
\end{equation}

(ii) for any integers $m_{1},m_{2}\geq 1$%
\[
\text{\textsc{Rev}}_{[m_{1}+m_{2}]}(X)\leq \text{\textsc{Rev}}_{[m_{1}]}(X)+%
\text{\textsc{Rev}}_{[m_{2}]}(X);
\]

(iii) the sequence $\frac{1}{m}$\textsc{Rev}$_{[m]}(X)$ is weakly decreasing
in $m,$ and thus, in particular, for every integer $m\geq 1$%
\begin{equation}
\text{\textsc{Rev}}_{[m]}(X)\leq m\cdot \text{\textsc{Rev}}_{[1]}(X);
\label{eq:m*rev1}
\end{equation}

(iv)%
\[
\text{\textsc{DRev}}(X)\leq (2^{k}-1)\cdot \text{\textsc{Rev}}%
_{[1]}(X)=(2^{k}-1)\cdot \text{\textsc{BRev}}(X). 
\]
\end{proposition}

\begin{proof}
(i) Let $\mu $ be any mechanism with a single menu entry, say $(g,t).$ If
the seller offers instead to sell the whole bundle at the same price $t,$
the buyer will surely buy whenever he did so in $\mu ,$ and the revenue can
only increase. Thus $R(\mu ;X)\leq $\textsc{BRev}$(X).$ Conversely, \textsc{%
BRev}$(X)$ is achieved by a single menu entry by Myerson's result (\ref%
{eq:one good}).

(ii) Let $\mu =(q,s)$ be any mechanism with menu $%
(g_{n},t_{n})_{n=1}^{m_{1}+m_{2}}$. For each menu entry $(g_{n},t_{n})$ let $%
\pi _{n}$ be the probability that it is chosen (when the valuation is $X$),
then the revenue from $\mu $ is $\sum_{n=1}^{m_{1}+m_{2}}\pi _{n}t_{n}.$ Let 
$\mu _{1}$ and $\mu _{2}$ be mechanisms with menus $%
(g_{n},t_{n})_{n=1}^{m_{1}}$ and $(g_{n},t_{n})_{n=m_{1}+1}^{m_{1}+m_{2}},$
respectively. The probability that $(g_{n},t_{n})$ for $n\leq m_{1}$ is
chosen is at least as large in $\mu _{1}$ as it is in $\mu $ (since every
valuation $x\in \mathbb{R}_{+}^{k}$ that prefers this menu item in $\mu $
continues to prefer it in $\mu _{1},$ and all $t_{n}$ are $\geq 0$ by NPT),
which implies that the revenue from $\mu _{1}$ is at least $%
\sum_{n=1}^{m_{1}}\pi _{n}t_{n}.$ A similar argument shows that the revenue
from $\mu _{2}$ is at least $\sum_{n=m_{1}+1}^{m_{1}+m_{2}}\pi _{n}t_{n}.$

(iii) Let $\mu =(q,s)$ be any mechanism with menu $(g_{n},t_{n})_{n=1}^{m};$
for each menu entry $(g_{n},t_{n})$ let $\pi _{n}$ be the probability that
it is chosen$.$ Without loss of generality order the menu entries so that
the sequence $\pi _{n}t_{n}$ is weakly decreasing. Let $m^{\prime }<m;$ the
mechanism $\mu ^{\prime }$ with menu $(g_{n},t_{n})_{n=1}^{m^{\prime }}$
yields as revenue at least $\sum_{n=1}^{m^{\prime }}\pi _{n}t_{n},$ which is
at least $(m^{\prime }/m)\sum_{n=1}^{m}\pi _{n}t_{n}$ (because $\pi _{n}t_{n}
$ is weakly decreasing). Thus $R(\mu ^{\prime };X)\geq (m^{\prime }/m)R(\mu
;X).$

(iv) A deterministic mechanism has menu size at most $2^{k}-1.$
\end{proof}

\bigskip

For small menu size $m$ the inequalites in (ii) and (iii) are tight, as the
example below shows that \textsc{Rev}$_{[m]}(X)=m\cdot $\textsc{Rev}$%
_{[1]}(X)$ for $m$ not exceeding the number of goods $k$. They remain
essentially tight for $m$ up to $2^{k}-1$ (by Theorem \ref{th:drev} below);
as for large $m,$ we will see that \textsc{Rev}$_{[m]}(X)$ can be as large as%
\footnote{%
It is convenient to use the standard O and $\Omega $ notations. For two
expressions $F$ and $G$ that depend on certain variables, we write $F=%
\mathrm{O}(G)$ if $\sup F/G<\infty ,$ and $F=\Omega (G)$ if $\inf F/G>0;$
i.e., there is a constant $0<c<\infty $ such that $F\leq cG,$ respectively $%
F\geq cG,$ for any values of the variables in the relevant range.} $\Omega
(m^{1/7})\cdot $\textsc{Rev}$_{[1]}(X)$ (see Theorem \ref{th:rev-m}).

\begin{example}
\label{ex:rm=m*r1}\noindent Let $1\leq m\leq k.$ Take a large\footnote{%
Theorem \ref{th:beta} below provides the tool to easily generate such
examples.} $H>0,$ and consider the following random valuation $X.$ For each $%
i=1,...,m,$ with a probability $\alpha _{i}$ that is proportional to $%
1/H^{i-1},$ good $i$ is valued at $H^{i-1}$ and all the other goods are
valued at $0;$ thus $\alpha _{i}=c/H^{i-1},$ where $%
c:=1/(1+1/H+...+1/H^{m-1}).$ Bundling yields a revenue of $1$ (because
setting the bundle price at $H^{i-1}$ yields a revenue of $%
c(1/H^{i-1}+...+1/H^{m-1}),$ which is maximal at $i=1,$ and the revenue
there is $1).$ Selling each good $i=1,...,m$ at price $H^{i-1}$ yields a
revenue of $c$ from each good; this is obtained at distinct valuations, and
so the mechanism consisting of these $m$ menu entries yields a revenue of $%
mc,$ which is close to $m$ for large $H.$
\end{example}

\subsection{Revenue Comparisons: \textsc{GFOR} and \textsc{MoB}\label%
{sus:gfor-mob}}

To evaluate how good mechanisms are we compare the revenue that they can
extract to two benchmarks, a \textquotedblleft high" one and a
\textquotedblleft low" one. The high benchmark is the \emph{optimal} revenue 
\textsc{Rev}, and the low benchmark is the \emph{basic} revenue \textsc{Rev}$%
_{[1]}=$\textsc{BRev}.\footnote{%
See Appendix \ref{ap:gamma} for a similar, but slightly less sharp, approach
where the basic revenue is taken to be the separate-selling revenue \textsc{%
SRev}.} As discussed in the Introduction, when the valuations of the goods
are correlated the former yields infinite gaps in most cases of interest,
and so the latter is needed to provide useful comparisons.

Formally, let $\mathbb{X}$ be a class of random valuations (e.g., $k$ goods,
two independent goods, and so on), and let $\mathcal{N}$ be a class of
mechanisms (e.g., separate mechanisms, deterministic mechanisms, and so on).

We define

\begin{itemize}
\item \textsc{GFOR}$(\mathcal{N};\mathbb{X}),$ the \emph{Guaranteed Fraction
of Optimal Revenue} (Hart and Nisan 2017), as the maximal fraction $\alpha $
such that, for any random valuation $X$ in $\mathbb{X},$ mechanisms in the
class $\mathcal{N}$ yield a revenue that is at least the fraction $\alpha $
of the optimal revenue; that is,\footnote{%
When taking the infimum we ignore the cases $0/0$ and $\infty /\infty $
(because the inequality $\mathcal{N}$-\textsc{Rev}$(X)\geq \alpha ~$\textsc{%
Rev}$(X)$ holds for any $\alpha $ in these cases). The same applies when
taking the supremum and, more generally, when dealing with any ratio of
revenues throughout the paper.}%
\[
\text{\textsc{GFOR}}(\mathcal{N};\mathbb{X}):=\inf_{X\in \mathbb{X}}\frac{%
\mathcal{N}\text{-\textsc{Rev}}(X)}{\text{\textsc{Rev}}(X)},
\]%
where $\mathcal{N}$-\textsc{Rev}$(X):=\sup_{\mu \in \mathcal{N}}R(\mu ;X)$
denotes the maximal revenue that can be obtained by any mechanism in the
class\footnote{\textsc{GFOR} is the reciprocal of the so-called
\textquotedblleft competitive ratio" used in the computer science
literature. While the two notions are clearly equivalent, using the optimal
revenue as the benchmark (i.e., 100\%) and measuring everything relative to
this basis---as \textsc{GFOR} does---seems to come more naturally. See the
remarks in Section 2.2 of Hart and Nisan (2017), which, in particular,
explain why ratios are used.} $\mathcal{N}.$

\item \textsc{MoB}$(\mathcal{N};\mathbb{X}),$ the \emph{Multiple of Basic
revenue}, as the minimal multiple $\beta $ such that, for any random
valuation $X$ in $\mathbb{X},$ mechanisms in the class $\mathcal{N}$ achieve
a revenue that is at most the multiple $\beta $ of the basic revenue; that
is,%
\[
\text{\textsc{MoB}}(\mathcal{N};\mathbb{X}):=\sup_{X\in \mathbb{X}}\frac{%
\mathcal{N}\text{-\textsc{Rev}}(X)}{\text{\textsc{Rev}}_{[1]}(X)}=\sup_{X\in 
\mathbb{X}}\frac{\mathcal{N}\text{-\textsc{Rev}}(X)}{\text{\textsc{BRev}}(X)}%
; 
\]%
when $\mathcal{N}$ consists of a single $k$-good mechanism $\mu $ and $%
\mathbb{X}$ is the class of all $k$-good random valuations we write \textsc{%
MoB}$(\mu )$ for short.
\end{itemize}

\noindent Thus, \textsc{MoB}$(\mathcal{N};\mathbb{X})$ is the highest
multiple of the basic revenue that may be achieved by the mechanisms in $%
\mathcal{N}$ for valuations in $\mathbb{X}.$

Putting $\alpha =$\textsc{GFOR}$(\mathcal{N};\mathbb{X})$ and $\beta =$%
\textsc{MoB}$(\mathcal{N};\mathbb{X}),$ we then have%
\[
\alpha \cdot \text{\textsc{Rev}}(X)\leq \mathcal{N}\text{-\textsc{Rev}}%
(X)\leq \beta \cdot \text{\textsc{Rev}}_{[1]}(X) 
\]%
for every random valuation $X$ in $\mathbb{X},$ and both bounds are tight:
i.e., for every $\alpha ^{\prime }>\alpha $ there is $X$ in $\mathbb{X}$
with $\alpha ^{\prime }\cdot $\textsc{Rev}$(X)>\mathcal{N}$-\textsc{Rev}$%
(X), $ and for every $\beta ^{\prime }<\beta $ there is $X$ in $\mathbb{X}$
with $\mathcal{N}$-\textsc{Rev}$(X)>\beta ^{\prime }\cdot $\textsc{Rev}$%
_{[1]}(X).$

\begin{remark}
\label{r:mob-le}The results of Proposition \ref{p:revm}(iii)--(iv) can be
thus restated as%
\begin{eqnarray*}
\text{\textsc{MoB}}(\text{\textsc{menu size}}\leq m;~k\text{ \emph{goods}})
&\leq &m\text{\ \ \ and} \\
\text{\textsc{MoB}}(\text{\textsc{deterministic}};~k\text{ \emph{goods}})
&\leq &2^{k}-1.
\end{eqnarray*}
\end{remark}

The following lemma provides simple but useful connections between \textsc{%
GFOR} and \textsc{MoB}.

\begin{lemma}
\label{l:gfor-mob}Let $\mathcal{M}$ be the class of all (IC and IR)
mechanisms, let $\mathcal{N\subset M}$, and let $\mathbb{X}$ be a class of
valuations. Then:

(i)%
\[
\text{\textsc{GFOR}}(\text{\textsc{bundled}};\mathbb{X})=\frac{1}{\text{%
\textsc{MoB}}(\mathcal{M};\mathbb{X})};\text{\ \ \ and} 
\]

(ii)%
\begin{eqnarray*}
\text{\textsc{GFOR}}(\mathcal{N};\mathbb{X}) &\leq &\text{\textsc{MoB}}(%
\mathcal{N};\mathbb{X})\cdot \text{\textsc{GFOR}}(\text{\textsc{bundled}};%
\mathbb{X}) \\
&=&\frac{\text{\textsc{MoB}}(\mathcal{N};\mathbb{X})}{\text{\textsc{MoB}}(%
\mathcal{M};\mathbb{X})}.
\end{eqnarray*}
\end{lemma}

\begin{proof}
(i) \textsc{GFOR}$($\textsc{bundled}$)=\inf_{X}$\textsc{BRev}$(X)/$\textsc{%
Rev}$(X)$ and \textsc{MoB}$(\mathcal{M})=\sup_{X}$\textsc{Rev}$(X)/$\textsc{%
BRev}$(X).$

(ii) $\mathcal{N}$-\textsc{Rev}$/$\textsc{Rev}$=(\mathcal{N}$-\textsc{Rev}$/$%
\textsc{BRev)~}$\cdot ~($\textsc{BRev}$/$\textsc{Rev}$).$
\end{proof}

\bigskip 

Thus, showing that there are mechanisms $\mu $ with arbitrarily large 
\textsc{MoB} proves that \textsc{GFOR}$($\textsc{bundled}$)=0$ (by (i)),
which then implies that\linebreak\ \textsc{GFOR}$($\textsc{menu size~}$\leq
m)=0$ for any fixed $m,$ and, in particular,\linebreak\ \textsc{GFOR}$($%
\textsc{deterministic}$)=0$ (by (ii) and Remark \ref{r:mob-le} above).

\section{Main Results\label{s:results}}

We now state formally the main results, first for the Guaranteed Fraction of
Optimal Revenue (\textsc{GFOR}), and then for the Multiple of Basic revenue (%
\textsc{MoB}), followed by an outline of the way in which these results are
proved.

\subsection{Results for \textsc{GFOR}\label{sus:results-gfor}}

The results here are, first, that \textsc{GFOR} equals $0$ for simple
mechanisms, including those with bounded menu size, and, second, that in the
case of bounded valuations \textsc{GFOR} becomes close to $1$ for an
appropriately large enough menu size. 

\begin{theorema}
\label{th:infinite gap}For $k\geq 2$ goods:

(i)%
\begin{eqnarray*}
\text{\textsc{GFOR}}(\text{\textsc{bundled}};\text{ }k\text{ goods}) &=&0; \\
\text{\textsc{GFOR}}(\text{\textsc{separate}};\text{ }k\text{ goods}) &=&0;
\\
\text{\textsc{GFOR}}(\text{\textsc{deterministic}};\text{ }k\text{ goods})
&=&0; \\
\text{\textsc{GFOR}}(\text{\textsc{menu}}\text{\textsc{\ size}}\leq m;\text{ 
}k\text{ goods}) &=&0
\end{eqnarray*}%
for every finite menu size $m\geq 1.$

(ii) For every $\varepsilon >0$ there exists a $k$-good random valuation $X$
with values in $[0,1]^{k}$ such that 
\[
\text{\textsc{DRev}}(X)<\varepsilon \cdot \text{\textsc{Rev}}(X). 
\]

(iii) There exists a $k$-good random valuation $X$ such that 
\[
\text{\textsc{DRev}}(X)=1~~~~\mathrm{and}~~~~\text{\textsc{Rev}}(X)=\infty . 
\]
\end{theorema}

As discussed in the Introduction, our contribution lies in the result for
the case of $k=2$ goods, as for $k\geq 3$ it follows from Briest et al.
(2014). In part (i), once we have the result that \textsc{GFOR} for \textsc{%
bundled }is $0$ all the other results immediately follow, because \textsc{Rev%
}$_{[m]}\leq m\cdot $\textsc{BRev} (by Proposition \ref{p:revm} and Lemma %
\ref{l:gfor-mob}(ii) above). Clearly, (ii) and (iii) each yield (i). If we
allow the valuations to be unbounded then we can get the fraction $%
\varepsilon $ in (ii) to go all the way down to $0,$ which is (iii). Clearly
(iii) implies (ii) (just truncate $X$ beyond a high enough value); the
construction that yields (ii) however is simpler and explicit. Part (ii)
(and thus (i)) is proved in Section \ref{s:better} and part (iii) in Section %
\ref{s:gap}.\footnote{%
Given the marginal distributions of the valuations of the two goods---which
determine the separate revenue---we obtain joint distributions for which the
revenue becomes arbitrarily \emph{large}; by contrast, Carroll (2017) looks
at the \emph{smallest} joint revenue for given marginals.}

Looking at the constructions used in the proof of Theorem \ref{th:infinite
gap}, one sees that the range of valuations (i.e., the support of $X$) is
exponential in the gap obtained; more precisely, if we restrict the values
of each good to being in a range that is bounded (from above as well as from
below, i.e., away from\footnote{%
Both bounds are needed, as rescaling $X$ rescales all revenues and so does
not affect the ratios between revenues.} $0),$ say, in the range $[L,H],$
then the gap becomes bounded by some constant power of $\log (H/L);$ see
Section \ref{s:bounded}, where we show that this exponential blowup in the
range is indeed needed. Our result is:

\begin{theorema}
\label{th:bounded}Let $k=2.$ There exists a constant $c<\infty $ such that
for every $0<L<H<\infty $ and $\varepsilon >0,$ 
\[
\text{\textsc{GFOR}}(\text{\textsc{menu size}}\leq m;\text{ }2\text{ goods
with values in\emph{\ }}[L,H]^{2})\geq 1-\varepsilon 
\]%
holds for every menu size $m$ that satisfies 
\[
m\geq \frac{c}{\varepsilon ^{5}}\log ^{2}\left( \frac{H}{L}\frac{1}{%
\varepsilon }\right) .
\]
\end{theorema}

This theorem is proved in Section \ref{s:bounded}. Again, contrast this
result with the unbounded range case: when the upper bound $H$ is infinite
(and $L>0)$ there is a valuation $X\ $with \textsc{Rev}$(X)=\infty $ while 
\textsc{Rev}$_{[m]}(X)\leq m$ for every finite $m$ (by\footnote{%
For the boundedness away from $0,$ see Remark \ref{r:X>0}.} Theorem \ref%
{th:infinite gap}(iii) and (\ref{eq:m*rev1})), and when the lower bound $L$
is zero (and $H$ is finite) for every finite $m$ there is a valuation $X$
with \textsc{Rev}$_{[m]}(X)/\text{\textsc{Rev}}(X)<m\varepsilon $ (by
Theorem \ref{th:infinite gap}(ii) and (\ref{eq:m*rev1})).

Thus arbitrarily good approximations of the optimal revenue can be obtained,
for two goods, by a menu size $m$ that is only \emph{polylogarithmic} in the
range size $H/L.$ This improves results obtainable by known techniques
(Hartline and Koluim 2005, Balcan et al.\emph{\ }2008, Briest et al\emph{.}
2015, and our Proposition \ref{th:additive-delta} below), which yield a 
\emph{polynomial} dependence on $H/L$ (i.e., $m\geq (H/L\varepsilon )^{ck}).$
Recently Dughmi, Han, and Nisan (2014) have extended the polylogarithmic
result to all $k$ (i.e., $m\geq (\log (H/(L\varepsilon ))/\varepsilon
)^{ck}),$ and shown that the exponential dependence on $k$ is necessary.

\subsection{Results for \textsc{MoB}\label{sus:results-mob}}

The results here show the relations between \textsc{MoB} and menu size
(polynomial), and, for deterministic and separate-selling mechanisms,
between \textsc{MoB} and the number of goods (exponential for the former and
linear for the latter).

\begin{theorema}
\label{th:rev-m}There exists a constant $c>0$ such that for every $k\geq 2$
and $m\geq 1$%
\[
cm^{1/7}\leq \text{\textsc{MoB}}(\text{\textsc{menu size}}\leq m;\text{ }k%
\text{ goods})\leq m.
\]
\end{theorema}

As discussed above, the right-hand side inequality, whose simple proof is in
Proposition \ref{p:revm}, says that the revenue may grow at most linearly in
the menu size; as for the left-hand side, which is obtained from our
construction in the proof of Theorem \ref{th:infinite gap}(iii) in Section %
\ref{s:gap}, it says that the revenue may grow at least polynomially in menu
size.\footnote{%
The increase is at a polynomial rate in $m$, and we do not think that the
constant of $1/7$ we obtain is tight. For larger values of $k$ the
construction in Briest \emph{et al.} (2015) implies a somewhat better
polynomial dependence on $m$. For $m$ that is at most exponential in $k$,
Theorem \ref{th:drev} below shows that the growth can be almost linear in $m.
$}

Returning to deterministic mechanisms, whose menu size is at most $2^{k}-1,$
we have the following.

\begin{theorema}
\label{th:drev}For every\footnote{%
We obtain in fact a lower bound that is somewhat better than $(2^{k}-1)/k$;
for large $k,$ it is close to twice as much. See Proposition \ref%
{p:drev/brev} and Remark \ref{r:d_k}.} $k\geq 2$:%
\begin{eqnarray}
\frac{2^{k}-1}{k} &\leq &\text{\textsc{MoB}}(\text{\textsc{deterministic}};%
\text{ }k\text{ goods})  \label{eq:drev-2^k} \\
&\leq &\text{\textsc{MoB}}(\text{\textsc{menu size}}\leq 2^{k}-1;\text{ }k%
\text{ goods})\leq 2^{k}-1.  \label{eq:2^k-1}
\end{eqnarray}
\end{theorema}

The upper bound (\ref{eq:2^k-1}) is given, again, by Proposition \ref{p:revm}%
; as for the lower bound (\ref{eq:drev-2^k}), which is proved using the
techniques of the proof of Theorem \ref{th:infinite gap}(iii) in Section \ref%
{s:gap}, it shows that the exponential-in-$k$ bound is essentially tight
(the factor $k$ being much smaller than $2^{k}-1$ for large $k).$ Note again
the contrast to the independent case, for which the bound is linear, rather
than exponential,\footnote{%
Proposition \ref{p:drev/brevsrev} in Appendix \ref{ap:gamma} below shows
that the same exponential-in-$k$ gap exists between deterministic mechanisms
and separate selling: there is $X$ such that \textsc{DRev}$(X)\geq
(2^{k}-1)/k\cdot $\textsc{SRev}$(X)$. This provides a rare \emph{doubly
exponential} contrast with the independent case in which \textsc{DRev}$%
(X)\leq c\log ^{2}k\cdot $\textsc{SRev}$(X)$ for some constant $c$ (by
Theorem C in Hart and Nisan 2017).} in $k$: Lemma 28 in Hart and Nisan
(2017) implies that for $k$ independent goods \textsc{DRev}$(X)\leq \text{%
\textsc{Rev}}(X)\leq ck\cdot $\textsc{Rev}$_{[1]}(X)$ for some $c>0,$ and
thus \textsc{MoB}$($\textsc{deterministic}; $k$ \emph{independent goods}$%
)\leq ck$.

The two inequalities in Theorem \ref{th:drev} say that the revenue that can
be extracted by deterministic mechanisms is, for large $k,$ of the same
order of magnitude as for arbitrary mechanisms with a menu of size $2^{k}-1.$
This suggests that the reason that deterministic mechanisms yield low
revenue (cf. Theorem \ref{th:infinite gap}) is \emph{not} that they are
deterministic, but rather that being deterministic limits their menu size
(to $2^{k}-1);$ any mechanism with that menu size will do just as badly.

Finally, we consider the maximal revenue \textsc{SRev} obtainable by selling
each good separately (at its one-good optimal price). We have

\begin{theorema}
\label{th:srev}For every $k\geq 2$: 
\[
\text{\textsc{MoB}}(\text{\textsc{separate}};\text{ }k\text{ goods})=k.
\]
\end{theorema}

This theorem is proved in Section \ref{s:additive menu}. Unlike in our
previous results, the bound here is the same as the one we have obtained for
independently distributed goods, and it is tight already in that case; see
Proposition 14(i) and Example 27 in Hart and Nisan (2017).

Now the mechanism that sells the $k$ goods separately has menu size $2^{k}-1$
(since the buyer may acquire any subset of the goods, and so there are $%
2^{k}-1$ possible outcomes), but its revenue may be at most $k$ times,
rather than $2^{k}-1$ times, the bundling revenue. Moreover, selling
separately seems intuitively to be much simpler than this exponential-in-$k$
menu-size measure suggests: one needs to determine only $k$ prices. All this
leads us to define a stronger notion of mechanism complexity, one that
assigns to separate selling its more natural complexity, namely, $k.$ This
new measure allows \textquotedblleft additive menus\textquotedblright\ in
which the buyer may choose not just single menu entries but also sets of
menu entries. We present this \emph{additive menu size\ }complexity\emph{\ }%
measure in Section \ref{s:additive menu}, and show that in fact our results
hold with respect to this stronger complexity measure as well.

\subsection{Outline of the Proofs\label{sus:proof-outline}}

We present now a short but hopefully useful outline of the proofs in the
following sections.

\begin{itemize}
\item In Section \ref{s:beta} we provide an explicit formula for \textsc{MoB}
of a mechanism, and construct random valuations where \textsc{MoB} is
(almost) attained (Theorem \ref{th:beta}).

\item In Section \ref{s:better} we construct mechanisms with an arbitrarily
large \textsc{MoB}, which shows that \textsc{MOB}($\mathcal{M})=\infty $ and
so \textsc{GFOR}$($\textsc{bundled}$)=0,$ thus proving Theorem \ref%
{th:infinite gap}(i) and (ii).

\item In Section \ref{s:gap} we construct a random valuation, for Theoreom %
\ref{th:infinite gap}(iii), with an infinite gap between the revenue from
simple mechanisms and the optimal revenue; we also prove the lower bound of
Theorem \ref{th:drev} for deterministic mechanisms.

\item In Section \ref{s:additive menu} we prove Theorem \ref{th:srev} for
the separate revenue, and then introduce and analyze the more refined
\textquotedblleft additive-menu-size" measure.

\item In Section \ref{s:bounded} we deal with valuations in bounded domains
and prove Theorem \ref{th:bounded}.
\end{itemize}

\section{The Multiple of Basic\textbf{\ Revenue (\textsc{MoB}})\textbf{\label%
{s:beta}}}

We start by providing a precise tool that measures how much better a
mechanism can be relative to bundling. It will then be used in the next
sections to construct random valuations together with corresponding
mechanisms that yield revenues that are arbitrarily higher than the bundling
revenue, and thus than any other simple revenue as well. Recall that for a
single $k$-good mechanism $\mu $ we write \textsc{MoB}$(\mu \mathcal{)}$ for
short for \textsc{MoB}$(\{\mu \};k$ goods$\mathcal{)}.$

\begin{theorem}
\label{th:beta}Let $\mu =(q,s)$ be a $k$-good mechanism. Then 
\[
\text{\textsc{MoB}}(\mu )=\int_{0}^{\infty }\frac{1}{v(t)}~\mathrm{d}t,
\]%
where for every $t>0$ we define\footnote{%
The $1$-norm $||x||_{1}=\sum_{i=1}^{k}|x_{i}|$ on $\mathbb{R}^{k}$ gives,
for nonnegative $x,$ the value $\sum_{i=1}^{k}x_{i}$ of the bundle of all
goods to the buyer of type $x$. The infimum of an empty set is taken to be $%
\infty $, and so $v(t)=\infty $ when $t$ is higher than any possible payment 
$s(x)$.}%
\[
v(t)%
%TCIMACRO{\TeXButton{:=}{{\;:=\;}}}%
%BeginExpansion
{\;:=\;}%
%EndExpansion
\inf \{||x||_{1}:x\in \mathbb{R}_{+}^{k}\text{ and }s(x)\geq t\}.
\]
\end{theorem}

Thus $v(t)$ is the minimal value of the bundle, $x_{1}+...+x_{k},$ among all
the valuations $x$ where the payment to the seller is at least $t.$
Geometrically, this says that the supporting hyperplane with normal $%
(1,...,1)$ to the set $\{x\in \mathbb{R}_{+}^{k}:s(x)\geq t\}$ is $%
x_{1}+...+x_{k}=v(t).$ The function $v$ is weakly increasing and satisfies $%
v(t)\geq t$ for every $t>0$ (because $\sum_{i}x_{i}\geq q(x)\cdot x\geq s(x)$
for every $x$ by IR); the function $1/v$ is nonnegative, weakly decreasing,
and vanishes beyond the maximal possible payment (i.e., for $t>\sup_{x}s(x)).
$ Its integral may well be zero or infinite, i.e., $0\leq $ \textsc{MoB}$%
(\mu )\leq \infty $ (with \textsc{MoB}$(\mu )=0$ only when $v(t)=\infty $
for every $t>0,$ which is the case only for the null mechanism with $s(x)=0$
for all $x$). When $\mu $ has a finite menu, say $\{(g_{n},t_{n})%
\}_{n=1}^{m},$ ordered so that the sequence $t_{n}$ is weakly increasing, we
have $v(t)=v(t_{n})$ for every $t_{n-1}<t\leq t_{n}$ (some of these
intervals may well be empty\footnote{%
If $v(t_{n})=v(t_{n+1})$ then we may eliminate $t_{n}$ altogether from the
sum, because\linebreak\ $%
(t_{n}-t_{n-1})/v(t_{n})+(t_{n+1}-t_{n})v(t_{n+1})=(t_{n+1}-t_{n-1})/v(t_{n+1})
$.}), and so%
\begin{equation}
\text{\textsc{MoB}}(\mu )=\sum_{n=1}^{m}\frac{t_{n}-t_{n-1}}{v(t_{n})}
\label{eq:beta-sum}
\end{equation}%
(computing the numbers $v(t_{n})$ amounts to solving $m$ linear programming
problems).

It may be instructive to compute \textsc{MoB}$(\mu )$ in a few examples with 
$k=2$ goods.

\begin{example}
\label{ex:mob}Let $\mu $ be given by the menu\footnote{%
We write a menu entry $(g,t)$ here as $g\cdot x-t;$ the payoff of the buyer
with valuation $x$ is thus $b(x)=\max
\{0,x_{1}-p_{1},x_{2}-2,x_{1}+x_{2}-4\}).$} $%
\{x_{1}-p_{1},x_{2}-2,x_{1}+x_{2}-4\},$ and allow $p_{1}$ to vary.

(i) When $p_{1}=1$ we have $(t_{1},t_{2},t_{3})=(1,2,4)$ and $%
(v(t_{1}),v(t_{2}),v(t_{3}))=(1,2,5)$ (attained, respectively, at the points 
$(1,0),$ $(0,2),$ and $(2,3);$ 
%TCIMACRO{\TeXButton{BeginFigure}{\begin{figure}[tb] \centering}}%
%BeginExpansion
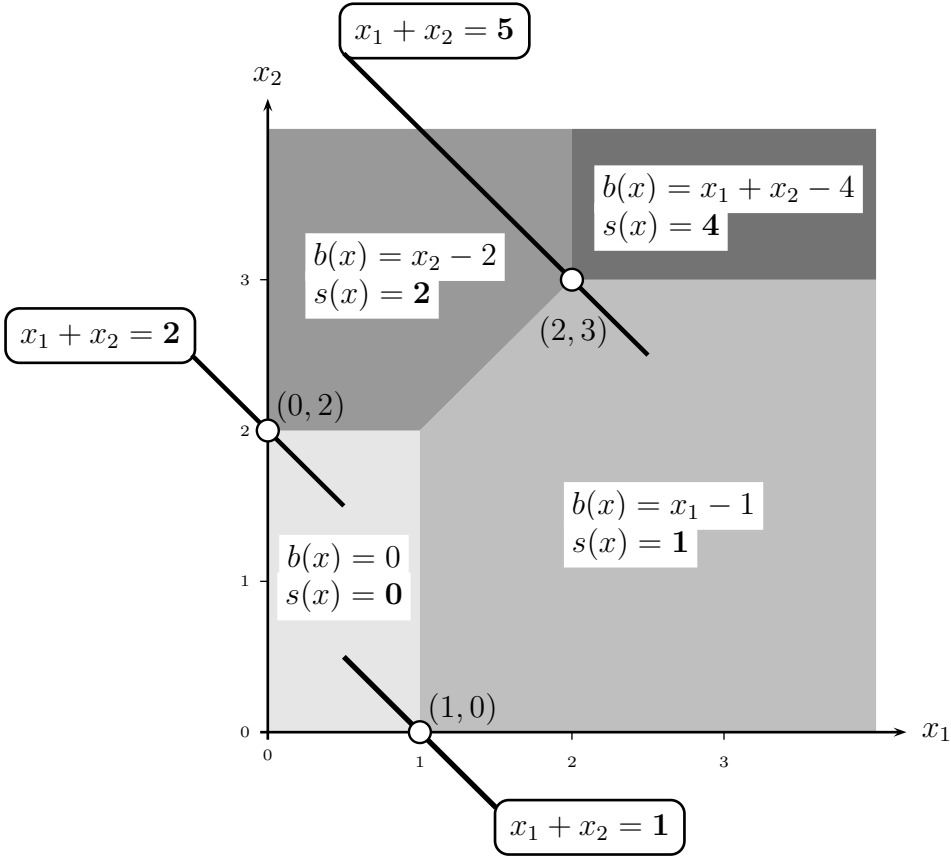
\begin{figure}[tb] \centering%
%EndExpansion
\begin{pspicture}(0,-2)(8,8)
    \psset{unit=2cm}

    %\psline[linewidth=1pt]{-}(1,0)(1,2)
    %\psline[linewidth=1pt]{-}(0,2)(1,2)
    %\psline[linewidth=1pt]{-}(1,2)(2,3)
    %\psline[linewidth=1pt]{-}(2,3)(2,4)
    %\psline[linewidth=1pt]{-}(2,3)(4,3)

    \psline[linewidth=0.5pt]{-}(2,-0.05)(2,0.05)
    \rput(2,-0.2){{\tiny $2$}}
    \psline[linewidth=0.5pt]{-}(1,-0.05)(1,0)
    \rput(1,-0.2){{\tiny $1$}}
    \psline[linewidth=0.5pt]{-}(3,-0.05)(3,0.05)
    \rput(3,-0.2){{\tiny $3$}}
    \psline[linewidth=0.5pt]{-}(-0.05,1)(0.05,1)
    \rput(-0.15,1){{\tiny $1$}}
    \psline[linewidth=0.5pt]{-}(-0.05,2)(0,2)
    \rput(-0.15,2){{\tiny $2$}}
    \psline[linewidth=0.5pt]{-}(-0.05,3)(0.05,3)
    \rput(-0.15,3){{\tiny $3$}}
    \rput(-0.15,0){{\tiny $0$}}
    \rput(0,-0.15){{\tiny $0$}}

    \newgray{gray9}{0.9}
    \newgray{gray8}{0.75}
    \newgray{gray7}{0.6}
    \newgray{gray6}{0.45}
    \pspolygon*[linecolor=gray9](0,0)(1,0)(1,2)(0,2)
    \pspolygon*[linecolor=gray8](1,0)(4,0)(4,3)(2,3)(1,2)
    \pspolygon*[linecolor=gray7](0,2)(1,2)(2,3)(2,4)(0,4)
    \pspolygon*[linecolor=gray6](2,3)(2,4)(4,4)(4,3)

    \psline[linewidth=1pt]{->}(-0.05,0)(4.2,0)
    \rput(4.4,0){$x_1$}
    \psline[linewidth=1pt]{->}(0,-0.05)(0,4.2)
    \rput(0,4.35){$x_2$}

    \psline[linewidth=2pt]{-}(0.5,0.5)(1.5,-0.5)
    \rput[l]{0}(1.485,-0.63){\psframebox[framesep=5pt,framearc=0.5,linewidth=1pt,linecolor=black]{$x_1+x_2={\mathbf 1}$}}
    \psline[linewidth=1.7pt]{-}(-0.5,2.5)(0.5,1.5)
    \rput[r]{0}(-0.47,2.63){\psframebox[framesep=5pt,framearc=0.5,linewidth=1pt,linecolor=black]{$x_1+x_2={\mathbf 2}$}}
    \psline[linewidth=1.7pt]{-}(2.5,2.5)(0.5,4.5)
    \rput[l]{0}(0.465,4.65){\psframebox[framesep=5pt,framearc=0.5,linewidth=1pt,linecolor=black]{$x_1+x_2={\mathbf 5}$}}

     %\psline[linewidth=1pt]{-}(0,2)(1,2)
     %\psline[linewidth=1pt]{-}(1,2)(2,3)

    \rput*(0.5,1.15){$b(x)=0$}
    \rput*(0.5,0.9){$s(x)={\mathbf 0}$}
    \rput*[l](2,1.5){$b(x)=x_1-1$}
    \rput*[l](2,1.25){$s(x)={\mathbf 1}$}
    \rput*[l](0.3,3.15){$b(x)=x_2-2$}
    \rput*[l](0.3,2.9){$s(x)={\mathbf 2}$}
    \rput*[l](2.2,3.6){$b(x)=x_1+x_2-4$}
    \rput*[l](2.2,3.35){$s(x)={\mathbf 4}$}

    %\psline[linewidth=4pt,linecolor=red]{->}(1,2.3)(2.1,2.6)
    %\psline[linewidth=4pt,linecolor=red]{->}(1.08,2.36)(1.92,2.64)
    \pscircle*[fillcolor=white,linecolor=white](1,0){0.07}
    \pscircle[linewidth=0.9pt](1,0){0.08}
    \pscircle*[fillcolor=white,linecolor=white](0,2){0.07}
    \pscircle[linewidth=0.9pt](0,2){0.08}
    \pscircle*[fillcolor=white,linecolor=white](2,3){0.07}
    \pscircle[linewidth=0.9pt](2,3){0.08}

    \rput[lb](1.05,0.05){{$(1,0)$}}
    \rput[lb](0.05,2.05){{$(0,2)$}}
    \rput[lb](1.78,2.55){{$(2,3)$}}
 \end{pspicture}\caption{The function $v$ in Example \ref{ex:mob}(i):
$v(1)=||(1,0)||_1=1, ~v(2)=||(0,2)||_1=2, ~v(4)=||(2,3)||_1=5$ \label{fig}}%
%TCIMACRO{\TeXButton{EndFigure}{\end{figure}}}%
%BeginExpansion
\end{figure}%
%EndExpansion
see Figure \ref{fig}). Therefore \textsc{MoB}$(\mu
)=(1-0)/1+(2-1)/2+(4-2)/5=19/10.$ As we will see in the proof of Theorem \ref%
{th:beta} below, \textsc{MoB}$(\mu )$ is attained for the random valuation $X
$ that takes the values $(1,0),$ $(0,2),$ and $(2,3)$ with probabilities $%
1/v(1)-1/v(2)=1/2,$ $1/v(2)-1/v(4)=3/10,$ and $1/v(4)=1/5,$ respectively;
indeed, \textsc{BRev}$(X)=\max \{(1+0)\cdot 1,(0+2)\cdot (1/2),(2+3)\cdot
(1/5)\}=1$ and\footnote{%
Assume without loss of generality that the buyer breaks ties in favor of the
seller (i.e., the mechanism $\mu $ is \textquotedblleft seller-favorable");
see Hart and Reny (2015).} $R(\mu ;X)=1\cdot (1/2)+2\cdot (3/10)+4\cdot
(1/5)=19/10$.

(ii) When $p_{1}=2$ we have $(t_{1},t_{2},t_{3})=(2,2,4)$ and $%
(v(t_{1}),v(t_{2}),v(t_{3}))=(2,2,4)$ (with $v(2)$ attained at $(2,0)$ and
also at $(0,2),$ and $v(4)$ at $(2,2)$). Therefore \textsc{MoB}$(\mu
)=(2-0)/2+(2-2)/2+(4-2)/4=3/2$.

(iii) When $p_{1}=5$ we have $(t_{1},t_{2},t_{3})=(2,4,5)$ and $%
(v(t_{1}),v(t_{2}),v(t_{3}))=(2,4,\infty )$ (with the first two attained at $%
(0,2)$ and $(2,2),$ and $v(5)$ infinite since $x_{1}-5$ is never chosen by
the buyer, as it is always strictly worse than $x_{1}+x_{2}-4).$ Therefore 
\textsc{MoB}$(\mu )=(2-0)/2+(4-2)/4+(5-4)/\infty =3/2$.
\end{example}

\begin{proof}[Proof of Theorem \protect\ref{th:beta}]
Put $\beta :=\int_{0}^{\infty }1/v(t)~\mathrm{d}t.$

(i) First, we show that 
\[
\frac{R(\mu ;X)}{\text{\textsc{BRev}}(X)}\leq \beta 
\]%
for every $k$-good random valuation $X.$ Indeed,%
\begin{eqnarray*}
R(\mu ;X) &=&\mathbb{E}\left[ s(X)\right] =\int_{0}^{\infty }\mathbb{P}\left[
s(X)\geq t\right] ~\mathrm{d}t\leq \int_{0}^{\infty }\mathbb{P}\left[
||X||_{1}\geq v(t)\right] ~\mathrm{d}t \\
&\leq &\int_{0}^{\infty }\frac{\text{\textsc{BRev}}(X)}{v(t)}~\mathrm{d}%
t=\beta \cdot \text{\textsc{BRev}}(X),
\end{eqnarray*}%
where we have used: $s(X)\geq 0$ by NPT; $s(X)\geq t$ implies $||X||_{1}\geq
v(t)$ by the definition of $v(t);$ and $u\cdot \mathbb{P}\left[
||X||_{1}\geq u\right] \leq ~$\textsc{BRev}$(X)$ for every $u>0.$

(ii) Second, we show that for every $\beta ^{\prime }<\beta $\ (which, when $%
\beta $ is infinite, is taken to mean any arbitrarily large $\beta ^{\prime }
$), there exists a $k$-good random valuation $X$ with $0<$\textsc{BRev}$%
(X)<\infty $\ and%
\begin{equation}
\frac{R(\mu ;X)}{\text{\textsc{BRev}}(X)}>\beta ^{\prime }.  \label{eq:beta'}
\end{equation}%
Indeed, the function $1/v(t)$ is weakly decreasing and nonnegative, and its
integral is $\beta ,$ and so there exist $0=t_{0}<t_{1}<...<t_{N}<t_{N+1}=%
\infty $ with $0=v(t_{0})<v(t_{1})<v(t_{2})<...<v(t_{N})<v(t_{N+1})=\infty $
such that%
\[
\beta ^{\prime \prime }:=\sum_{n=1}^{N}\frac{t_{n}-t_{n-1}}{v(t_{n})}>\beta
^{\prime }.
\]%
Let $\varepsilon >0$ be small enough so that $\beta ^{\prime \prime
}>(1+\varepsilon )\beta ^{\prime }$ and $v(t_{n+1})>(1+\varepsilon )v(t_{n})$
for all $1\leq n\leq N.$ By the definition of $v$ we can choose for every $%
1\leq n\leq N$ a point\footnote{%
Subscripts $n,m,$ and $j$ are used for sequences, whereas $i$ is used
exclusively for coordinates; thus $x_{n}$ is a vector in $\mathbb{R}_{+}^{k},
$ and $x_{i}$ is the $i$-th coordinate of $x.$} $x_{n}\in \mathbb{R}_{+}^{k}$
such that $s(x_{n})\geq t_{n}$ and $v(t_{n})\leq
||x_{n}||_{1}<(1+\varepsilon )v(t_{n})$; then%
\begin{equation}
\sum_{n=1}^{N}\frac{t_{n}-t_{n-1}}{||x_{n}||_{1}}>\sum_{n=1}^{N}\frac{%
t_{n}-t_{n-1}}{v(t_{n})(1+\varepsilon )}=\frac{\beta ^{\prime \prime }}{%
1+\varepsilon }>\beta ^{\prime }.  \label{eq:sum-beta'}
\end{equation}

Put $\xi _{n}:=||x_{n}||_{1};$ the sequence $\xi _{n}$ is strictly
increasing (because $(1+\varepsilon )v(t_{n})<v(t_{n+1})$) and $\xi _{1}>0$
(because $v(t_{1})>0).$ Let $X$ be a random variable with support $%
\{x_{1},...,x_{N}\}$ and distribution $\mathbb{P}\left[ X=x_{n}\right] =\xi
_{1}/\xi _{n}-\xi _{1}/\xi _{n+1}$ for every $1\leq n\leq N,$ where we put $%
\xi _{N+1}:=\infty ;$ thus $\mathbb{P}\left[ X\in \{x_{n},...,x_{N}\}\right]
=\xi _{1}/\xi _{n}$ for every\footnote{%
Since the payment $s(x_{n})$ increases with $n,$ we want to put as much
probability as possible on points $x_{n}$ with high $n,$ subject to the
constraint that the bundled revenue is kept fixed, specifically, equal to $%
\xi _{1}=||x_{1}||_{1};$ for illustration see the random valuation $X$ in
Example \ref{ex:mob}(i) above. } $n\geq 1.$

To compute \textsc{BRev}$(X),$ we need to consider only the bundle prices $%
\xi _{n}$ for $1\leq n\leq N$ (these are the possible values of $%
\sum_{i}X_{i}=||X||_{1}),$ for which we have%
\[
\xi _{n}\cdot \mathbb{P}\left[ ||X||_{1}\geq \xi _{n}\right] =\xi _{n}\cdot 
\mathbb{P}\left[ X\in \{x_{n},...,x_{N}\}\right] =\xi _{n}\cdot \frac{\xi
_{1}}{\xi _{n}}=\xi _{1}, 
\]%
and so 
\begin{equation}
\text{\textsc{BRev}}(X)=\xi _{1}.  \label{eq:brev=xi1}
\end{equation}

Finally, the revenue $R(\mu ;X)$ that $\mu $ extracts from $X$ is%
\begin{eqnarray}
R(\mu ;X) &\geq &\sum_{n=1}^{N}s(x_{n})\mathbb{P}\left[ X=x_{n}\right] \geq
\sum_{n=1}^{N}t_{n}\left( \frac{\xi _{1}}{\xi _{n}}-\frac{\xi _{1}}{\xi
_{n+1}}\right)  \label{eq:R} \\
&=&\sum_{n=1}^{N}(t_{n}-t_{n-1})\frac{\xi _{1}}{\xi _{n}}>\xi _{1}\beta
^{\prime }=\beta ^{\prime }\cdot \text{\textsc{BRev}}(X)  \nonumber
\end{eqnarray}%
(use $\xi _{N+1}=\infty ,$ (\ref{eq:sum-beta'}), and (\ref{eq:brev=xi1})).
\end{proof}

\begin{remark}
\label{r:mob=int(1/v)}\emph{(a) }In the proof of part (ii) above: for any $%
m<N$ let $\mu _{m}$ be obtained by restricting the menu of $\mu $ to the
entries chosen by $x_{1},...,x_{m}$ in $\mu $ (with ties broken the same way
as in $\mu $ for $x_{1},...,x_{m},$ and arbitrarily otherwise)$.$\footnote{%
Formally, $\mu _{m}=(q_{m},s_{m})$ satisfies $(q_{m}(x),s_{m}(x))=(q(x),s(x))
$ for $x\in \{x_{1},...,x_{m}\}$ and $(q_{m}(x),s_{m}(x))\in \arg
\max_{1\leq n\leq m}(q(x_{n})\cdot x-s(x_{n}))$ otherwise.} The computation
of $R(\mu _{m};X)$ is the same as in (\ref{eq:R}), but the sum is now going
up only to $m$ instead of $N,$ and thus there is a final term of $t_{m}(\xi
_{1}/\xi _{m+1})$ that needs to be subtracted; this gives%
\begin{equation}
R(\mu _{m};X)>\left( \sum_{n=1}^{m}\frac{t_{n}-t_{n-1}}{\xi _{n}}-\frac{t_{m}%
}{\xi _{m+1}}\right) \cdot \text{\textsc{BRev}}(X)  \label{eq:mu-m}
\end{equation}%
(recall (\ref{eq:brev=xi1})). This result will be used in Proposition \ref%
{p:gap} below.

\emph{(b) }The random valuation $X$ that we have constructed in part (ii) of
the proof has finite support, and is thus bounded from above; one may
therefore rescale it (which does not affect the ratio of revenues)\ so that
it takes values in, say, $[0,1]^{k}.$

\emph{(c)} If the mechanism $\mu $ has a finite menu of size $m$ then $v(t)$
can take at most $m$ distinct values, and so $N\leq m$ and the support of
the resulting $X$ is of size at most $m.$

\emph{(d)} If the mechanism $\mu $ has a finite menu of size $m$ then 
\textsc{MoB}$(\mu )\leq m$ (because $v(t)\geq t$ implies that each term in
the sum (\ref{eq:beta-sum}) is $\leq 1).$ This is the linear-in-menu-size
bound of Proposition \ref{p:revm}(iii); Example \ref{ex:rm=m*r1} in Section %
\ref{sus:rev-m} above is obtained by making each term close to $1.$
\end{remark}

In Appendix \ref{ap:gamma} we will provide a similar analysis with the
separate revenue instead of the bundling revenue; it will use the $\infty $%
-norm instead of the $1$-norm.

\section{The Guaranteed Fraction of Optimal Revenue (\textsc{GFOR})\textsc{\ 
}\label{s:better}}

Based on the result of the previous section we can now construct mechanisms
whose revenues may be arbitrarily higher than the bundling revenue, which
yields the \textsc{GFOR}$=0$ result.

\begin{proposition}
\label{p:lnM}Let $k=2.$ For every finite $m\geq 1$ there exists a two-good
mechanism $\mu $ with a menu of size $m$ such that 
\[
\text{\textsc{MOB}}(\mu )>\frac{1}{2}\ln m-1. 
\]
\end{proposition}

\begin{proof}
Let $m=(N+1)^{2}-1$ where $N\geq 2$ is an integer. Let $g_{0},g_{1},...,g_{m}
$ be the $m+1=(N+1)^{2}$ points of the $1/N$-grid of $[0,1]^{2}$ arranged in
the lexicographic order, i.e., in order of increasing first coordinate, and,
for equal first coordinate, in order of increasing second coordinate (thus $%
g_{0}=(0,0)$ and $g_{m}=(1,1)).$

For each $n\geq 1,$ by writing the vector $g_{n}$ as $%
g_{n}=(i_{1}^{{}}/N,i_{2}^{{}}/N)$ with $i_{1}\equiv i_{1}^{(n)}$ and $%
i_{2}\equiv i_{2}^{(n)}$ integers between $0$ and $N,$ we define $%
y_{n}:=(N+1-i_{2},1).$ We claim that for every $0\leq j<n$ we have%
\begin{equation}
(g_{n}-g_{j})\cdot y_{n}\geq \frac{1}{N}.  \label{eq:gap_1/N}
\end{equation}%
Indeed, let $g_{j}=(\ell _{1}/N,\ell _{2}/N).$ Now $j<n$ implies either (i) $%
i_{1}=\ell _{1}$ and $i_{2}\geq \ell _{2}+1$, in which case $%
(g_{n}-g_{j})\cdot y_{n}=i_{2}/N-\ell _{2}/N\geq 1/N$, or (ii) $i_{1}\geq
\ell _{1}+1,$ in which case $(g_{n}-g_{j})\cdot y_{n}=(i_{1}/N-\ell
_{1}/N)(N+1-i_{2})+(i_{2}/N-\ell _{2}/N)\geq 1/N$ because $i_{1}-\ell
_{1}\geq 1$ and $i_{2}-\ell _{2}\geq 0-N=-N.$

Let $t_{n}:=N^{n-1}$ and $x_{n}:=N^{n}y_{n},$ and consider the mechanism $%
\mu =(q,s)$ with menu $\{(g_{n},t_{n})\}_{n=1}^{m}$ that is
\textquotedblleft seller-favorable"; i.e., when indifferent, the buyer
chooses the outcome with the highest payment (that is, ties are broken in
favor of the seller; see Hart and Reny 2015). For every $0\leq j<n$ we have%
\[
g_{n}\cdot x_{n}-g_{j}\cdot x_{n}=N^{n}(g_{n}-g_{j})\cdot y_{n}\geq
N^{n-1}=t_{n}\geq t_{n}-t_{j}, 
\]%
and so $g_{n}\cdot x_{n}-t_{n}\geq g_{j}\cdot x_{n}-t_{j}$. Therefore a
buyer of type $x_{n}$ will not choose any menu entry $(g_{j},t_{j})$ with $%
j<n$ (by seller-favorability when there is indifference, because $%
t_{j}<t_{n})$, and so $s(x_{n})$ is one of $\{t_{n},t_{n+1},...,t_{m}\},$
which implies that $s(x_{n})\geq t_{n}$. Thus $v(t_{n})\leq
||x_{n}||_{1}=N^{n}(N+2-i_{2}^{(n)}),$ and so%
\begin{eqnarray*}
\text{\textsc{MoB}}(\mu ) &=&\sum_{n=1}^{m}\frac{t_{n}-t_{n-1}}{v(t_{n})}%
\geq \sum_{n=1}^{m}\frac{N^{n-1}-N^{n-2}}{N^{n}(N+2-i_{2}^{(n)})} \\
&\geq &\sum_{i_{1}=1}^{N}\sum_{i_{2}=1}^{N}\frac{1}{N}\frac{1}{N+2-i_{2}}%
=\sum_{\ell =2}^{N+1}\frac{1}{\ell } \\
&>&\ln (N+2)-1>\frac{1}{2}\ln m-1
\end{eqnarray*}%
(in the second line we have dropped the terms with $i_{1}=0$ or $i_{2}=0).$
\end{proof}

\bigskip

Thus \textsc{MoB}$($\textsc{menu size} $\leq m)$ is at least of the order of 
$\log m;$ in the next section we will improve this lower bound and show that
it is polynomial in $m.$ From Proposition \ref{p:lnM} we immediately get
parts (i) and (ii) of Theorem \ref{th:infinite gap}.

\bigskip

\begin{proof}[Proof of Theorem \protect\ref{th:infinite gap}(i) and (ii)]
We prove this for $k=2$ goods; for $k>2$ goods we take the two-good random
valuation and append $k-2$ goods with constant valuation $0,$ which does not
affect any of the revenues$.$

(i) We have \textsc{MoB}$(\mathcal{M}$; $2\mathcal{\ }$\emph{goods}$\mathcal{%
)}=\sup_{\mu }$\textsc{MoB}$(\mu \mathcal{)=\infty }$ by Proposition \ref%
{p:lnM}, and so \textsc{GFOR}$($\textsc{bundled}; $2\mathcal{\ }$\emph{goods}%
$)=1/$\textsc{MoB}$(\mathcal{M}$; $2\ $\emph{goods}$\mathcal{)}=0$ (see
Lemma \ref{l:gfor-mob}(ii) in Section \ref{sus:gfor-mob}).

(ii) For every finite $m\geq 1,$ let $\mu $ be the mechanism given by
Proposition \ref{p:lnM}, and then let $X$ be a random valuation in $%
[0,1]^{2} $ with support of size $m$, as constructed by Theorem \ref{th:beta}
(see Remark \ref{r:mob=int(1/v)}(b) and (c)), that satisfies 
\begin{equation}
\frac{\text{\textsc{Rev}}(X)}{\text{\textsc{BRev}}(X)}\geq \frac{R(\mu ;X)}{%
\text{\textsc{BRev}}(X)}>\frac{1}{2}\ln m-1.  \label{eq:lnM}
\end{equation}

An explicit random valuation $X$ that satisfies (\ref{eq:lnM}) is easily
obtained from the proof of Proposition \ref{p:lnM}. Take $%
x_{n}=N^{n}y_{n}=N^{n}(N+1-i_{2}^{n},1),$ put $\xi _{n}:=||x_{n}||_{1},$ and
let $X$ have support $\{x_{1},...,x_{m}\}$ and distribution $\mathbb{P}\left[
X=x_{n}\right] =\xi _{1}/\xi _{n}-\xi _{1}/\xi _{n+1}$ for every $1\leq
n\leq m.$ Then \textsc{BRev}$(X)=\xi _{1}$ and \textsc{Rev}$(X)\geq R(\mu
;X)>\xi _{1}((1/2)\ln m-1)$ (cf. the proof of Theorem \ref{th:beta}). To get
the valuations in $[0,1]^{2}$ one just needs to rescale: divide everything
by $N^{m}.$ Taking $m$ large enough so that $(1/2)\ln m-1>1/(3\varepsilon )$
then yields (use Proposition \ref{p:revm}(iv)) \textsc{DRev}$(X)\leq 3\cdot $%
\textsc{BRev}$(X)<\varepsilon \cdot $\textsc{Rev}$(X)$.
\end{proof}

\begin{remark}
\label{r:pathologic}\emph{A}ny random valuation $X^{\prime }$ that is close
to the above random valuation $X$ will yield a similar gap between the
optimal revenue and the simple revenues.\footnote{%
For formal revenue continuity results, see Hart and Reny (2017, Appendix A).}
The same applies to all our constructions, and so none of our results is
knife-edge or pathological.
\end{remark}

\section{A General Construction\label{s:gap}}

We now generalize the construction of the previous section, and obtain a
mechanism $\mu $ with infinite \textsc{MoB}$,$ together with a corresponding
random valuation $X$ for which the optimal revenue is infinite, whereas all
its simple revenues---bundled, separate, deterministic, finite-menu---are
bounded; this proves Theorem \ref{th:infinite gap}(iii). Proposition \ref%
{p:gap} below will turn out to be useful also for evaluating \textsc{MoB} of
deterministic mechanisms, thereby proving Theorem \ref{th:drev}.

\begin{proposition}
\label{p:gap}Let $(g_{n})_{n=0}^{N}$ be a finite or countably infinite
sequence in $[0,1]^{k}$ starting with $g_{0}=(0,...,0),$ and let $%
(y_{n})_{n=1}^{N}$ be a sequence of points in $\mathbb{R}_{+}^{k}$ such that%
\[
\mathrm{gap}_{n}:=\min_{0\leq j<n}(g_{n}-g_{j})\cdot y_{n}>0 
\]%
for all $n\geq 1.$ Then for every $\varepsilon >0$ there exist a sequence $%
(t_{n})_{n=1}^{N}$ of positive real numbers, a $k$-good mechanism $\mu $
with menu $\{(g_{n},t_{n})\}_{n=1}^{N},$ and a $k$-good random valuation $X$
with $0<$\textsc{BRev}$(X)<\infty $ such that%
\begin{eqnarray}
\text{\textsc{MoB}}(\mu ) &\geq &(1-\varepsilon )\sum_{n=1}^{N}\frac{\mathrm{%
gap}_{n}}{||y_{n}||_{1}},  \nonumber \\
\frac{\text{\textsc{Rev}}(X)}{\text{\textsc{BRev}}(X)}\geq \frac{R(\mu ;X)}{%
\text{\textsc{BRev}}(X)} &\geq &(1-\varepsilon )\sum_{n=1}^{N}\frac{\mathrm{%
gap}_{n}}{||y_{n}||_{1}},\text{\ \ \ and}  \label{eq:sum-gaps} \\
\frac{\text{\textsc{Rev}}_{[m]}(X)}{\text{\textsc{BRev}}(X)}\geq \frac{R(\mu
_{m};X)}{\text{\textsc{BRev}}(X)} &\geq &(1-\varepsilon )\sum_{n=1}^{m}\frac{%
\mathrm{gap}_{n}}{||y_{n}||_{1}}-\varepsilon  \label{eq:sum-gaps-m}
\end{eqnarray}%
for every finite $1\leq m<N,$ where $\mu _{m}$ denotes the mechanism
obtained by restricting $\mu $ to its first $m$ menu entries $%
\{(g_{n},t_{n})\}_{n=1}^{m}$.
\end{proposition}

\begin{proof}
Let $x_{n}:=(t_{n}/\mathrm{gap}_{n})y_{n}$ where the sequence of positive
numbers $(t_{n})_{n\geq 1}$ increases fast enough so that the sequence $\xi
_{n}:=||x_{n}||_{1}=t_{n}||y_{n}||_{1}/\mathrm{gap}_{n}$ is increasing and $%
t_{n+1}/t_{n}\geq 1/\varepsilon $ for all $n\geq 1$. We have $\xi _{n}\geq
t_{n}$ (because $\mathrm{gap}_{n}\leq g_{n}\cdot y_{n}\leq ||y_{n}||_{1})$
and thus, when $N$ is infinite, $(t_{n})_{n}$ and $(\xi _{n})_{n}$ both
increase to infinity; when $N$ is finite, we put $t_{N+1}=\xi _{N+1}=\infty .
$ For every $0\leq j<n$,%
\[
g_{n}\cdot x_{n}-g_{j}\cdot x_{n}=\frac{t_{n}}{\mathrm{gap}_{n}}%
(g_{n}-g_{j})\cdot y_{n}\geq t_{n}\geq t_{n}-t_{j}
\]%
(for $j=0$ put as usual $t_{0}=0).$ Thus, in the seller-favorable mechanism $%
\mu =(q,s)$ with menu $\{(g_{n},t_{n})\}_{n=1}^{N},$ the buyer of type $x_{n}
$ prefers the menu entry $(g_{n},t_{n})$ to any entry $(g_{j},t_{j})$ with $%
0\leq j<n.$ Therefore $s(x_{n})\geq t_{n},$ and so $v(t_{n})\leq
||x_{n}||_{1}=\xi _{n},$ and we get%
\begin{eqnarray}
\text{\textsc{MoB}}(\mu ) &=&\sum_{n=1}^{N}\frac{t_{n}-t_{n-1}}{v(t_{n})}%
\geq \sum_{n=1}^{N}\frac{t_{n}-t_{n-1}}{\xi _{n}}  \nonumber \\
&=&\sum_{n=1}^{N}\frac{t_{n}-t_{n-1}}{t_{n}}\frac{\mathrm{gap}_{n}}{%
||y_{n}||_{1}}\geq (1-\varepsilon )\sum_{n=1}^{N}\frac{\mathrm{gap}_{n}}{%
||y_{n}||_{1}}  \label{eq:Gap}
\end{eqnarray}%
(the final inequality follows from $t_{n-1}/t_{n}\leq \varepsilon ).$ As in
the proof of Theorem \ref{th:beta}, let $X$ take the value $x_{n}$ with
probability $\xi _{1}/\xi _{n}-\xi _{1}/\xi _{n+1}$, then $R(\mu ;X)\geq \xi
_{1}\cdot \sum_{n=1}^{N}(t_{n}-t_{n-1})/\xi _{n}$ and \textsc{BRev}$(X)\leq
\xi _{1},$ which implies (\ref{eq:sum-gaps}) (use (\ref{eq:Gap})); to get (%
\ref{eq:sum-gaps-m}) for a finite $m<N,$ use (\ref{eq:mu-m}) and $\xi
_{m+1}\geq t_{m+1}\geq t_{m}/\varepsilon $.
\end{proof}

\begin{remark}
\label{r:X>0}The random valuation $X$ that we have constructed in
Proposition \ref{p:gap} is bounded away from zero: $||X||_{1}\geq
||x_{1}||_{1}=\xi _{1}>0.$
\end{remark}

Before showing how to obtain the infinite separation of Theorem \ref%
{th:infinite gap}(iii), we use Proposition \ref{p:gap} for deterministic
mechanisms, proving the lower bound on \textsc{MoB}$($\textsc{deterministic}$%
)$ of Theorem \ref{th:drev} (recall that the upper bound of $2^{k}-1$ is
immediate; see Proposition \ref{p:revm}(iv)).

\begin{proposition}
\label{p:drev/brev}For every $k\geq 2,$%
\begin{equation}
\text{\textsc{MoB}}(\text{\textsc{deterministic}};\text{ }k\text{ goods}%
)\geq \sum_{\ell =1}^{k}\frac{1}{\ell }\binom{k}{\ell }>\frac{2^{k}-1}{k}.
\label{eq:binomial-sum}
\end{equation}
\end{proposition}

\begin{proof}
Let $I_{0},I_{1},I_{2},...,I_{2^{k}-1}$ be the $2^{k}$ subsets of $%
\{1,\ldots ,k\}$ ordered in weakly increasing size (i.e., $|I_{n}|\geq
|I_{n-1}|$ for all $n),$ and let $g_{n}$ be the indicator vector of $I_{n}$
(i.e., the $i$-th coordinate of $g_{n}$ is $1$ for $i\in I_{n}$ and $0$ for $%
i\not\in I_{n}$). Take $y_{n}=g_{n}$ (thus $||y_{n}||_{1}=|I_{n}|);$ then
for $0\leq j<n$ we have $g_{j}\cdot g_{n}=|I_{j}\cap
I_{n}|<|I_{n}|=g_{n}\cdot g_{n}$ (the strict inequality holds because
otherwise $I_{n}$ would be a subset of $I_{j},$ contradicting $|I_{j}|\leq
|I_{n}|$ and $j\neq n),$ and thus $\mathrm{gap}_{n}\geq 1$ (in fact, $%
\mathrm{gap}_{n}=1$: take $I_{j}$ to be a subset of $I_{n}$ with one less
element). Thus%
\[
\sum_{n=1}^{2^{k}-1}\frac{\mathrm{gap}_{n}}{||y_{n}||_{1}}\geq
\sum_{n=1}^{2^{k}-1}\frac{1}{|I_{n}|}=\sum_{\ell =1}^{k}\frac{1}{\ell }%
\binom{k}{\ell },
\]%
and we use Proposition \ref{p:gap}. Replacing each $1/\ell $ with the lower $%
1/k$ yields the final inequality.
\end{proof}

\begin{remark}
\label{r:d_k}Let $d_{k}$ denote the binomial sum in (\ref{eq:binomial-sum}).

\emph{(a) }A better lower bound on $d_{k}$, easily obtained by replacing
each $1/\ell $ with the lower $1/(\ell +1),$ is\footnote{%
The standard notation $f(k)\sim g(k)$ means that $f(k)/g(k)\rightarrow 1$ as 
$k\rightarrow \infty .$} $d_{k}\geq (2^{k+1}-k-2)/(k+1)\sim 2\cdot
(2^{k}-1)/k.$

\emph{(b) }For large $k$ most of the mass of the binomial coefficients,
whose sum is $2^{k}-1,$ is at those $\ell $ that are close to $k/2,$ and so $%
d_{k}\sim 1/(k/2)\cdot (2^{k}-1)=2\cdot (2^{k}-1)/k$ (formally, use a
standard large deviation inequality; in (a) above we got this estimate only
as a lower bound on $d_{k}$).

\emph{(c) }For $k=2$ goods we have $d_{2}=\binom{2}{1}/1+\binom{2}{2}/2=5/2,$
which turns out to be the exact value of \textsc{MoB}; see Proposition \ref%
{p:5/2} in Appendix \ref{ap:beta} (proved by using, again, Theorem \ref%
{th:beta})$.$

\emph{(d)} Proposition \ref{p:drev/brevsrev} in Appendix \ref{ap:gamma}
shows that the same lower bound of $(2^{k}-1)/k$ also holds relative to the
separate (instead of the bundling) revenue, and even relative to the maximum
of the two revenues.
\end{remark}

We now construct, already for two goods, an infinite sequence of points for
which the appropriate sum of gaps in Proposition \ref{p:gap} is infinite.

\begin{proposition}
\label{qn-seq} There exists an infinite sequence of points $%
(g_{n})_{n=1}^{\infty }$ in $[0,1]^{2}$ with $||g_{n}||_{2}\leq 1$ such that
taking $y_{n}=g_{n}$ for all $n$ we have $\mathrm{gap}_{n}=\Omega (n^{-6/7})$%
.
\end{proposition}

\begin{proof}
The sequence of points that we build is composed of a sequence of
\textquotedblleft shells,\textquotedblright\ each containing multiple
points. The shells get closer and closer to each other, approaching the unit
sphere as the shell, $N$, goes to infinity: all the points $g_{n}$ in the $N$%
-th shell are of length $||g_{n}||_{2}=\sum_{\ell =1}^{N}\ell ^{-3/2}/\alpha 
$, where $\alpha =\sum_{\ell =1}^{\infty }\ell ^{-3/2}$ (which indeed
converges; thus $||g_{n}||_{2}$ approaches $1$ as $n$ increases), and each
shell $N$ contains $N^{3/4}$ different points in it so that the angle
between any two of them is at least $\Omega (N^{-3/4})$.

We now estimate $g_{n}\cdot g_{j}=||g_{n}||_{2}\cdot ||g_{j}||_{2}\cdot \cos
(\theta ),$ where $\theta $ denotes the angle between $g_{n}$ and $g_{j}$.
Let $N$ be $g_{n}$'s shell. For $j<n$ there are two possibilities: either $%
g_{j}$ is in the same shell, $N,$ as $g_{n}$ or it is in a smaller shell $%
N^{\prime }<N.$ In the first case we have $\theta \geq \Omega (N^{-3/4})$
and thus $\cos (\theta )\leq 1-\Omega (N^{-3/2})$ (because $\cos
(x)=1-x^{2}/2+x^{4}/24-\ldots $) and since $||g_{n}||_{2}=\Theta (1)$ we
have $g_{n}\cdot g_{n}-g_{n}\cdot g_{j}\geq \Omega (N^{-3/2})$. In the
second case, $||g_{n}||_{2}-||g_{j}||_{2}=\sum_{\ell =N^{\prime }+1}^{N}\ell
^{-3/2}/\alpha \geq N^{-3/2}/\alpha $, and so again since $%
||g_{n}||_{2}=\Theta (1)$ we have $g_{n}\cdot g_{n}-g_{n}\cdot g_{j}\geq
\Omega (N^{-3/2})$. Thus for any point $g_{n}$ in the $N$-th shell we have $%
\mathrm{gap}_{n}=\Omega (N^{-3/2})$. Since the first $N$ shells together
contain $\sum_{\ell =1}^{N}\ell ^{3/4}=\Theta (N^{7/4})$ points, we have $%
n=\Theta (N^{7/4})$ and thus \textrm{gap}$_{n}=\Omega (N^{-3/2})=\Omega
(n^{-6/7})$.
\end{proof}

\bigskip

This directly implies Theorem \ref{th:infinite gap}(iii), i.e., the infinite
separation between the optimal revenue and the deterministic revenue, and
also the lower bound in Theorem \ref{th:rev-m}, i.e., the revenue may
increase polynomially in the menu size.

\bigskip

\begin{proof}[Proof of Theorems \protect\ref{th:infinite gap}(iii) and 
\protect\ref{th:rev-m}]
For $k=2$ the infinite sequence of points $(g_{n})_{n=1}^{\infty }$
constructed in Proposition \ref{qn-seq}, together with $y_{n}=g_{n}$ for all 
$n,$ satisfies $\sum_{n=1}^{m}\mathrm{gap}_{n}/||g_{n}||_{1}\geq
\sum_{n=1}^{m}\mathrm{gap}_{n}/\sqrt{2}\geq \Omega (\sum_{n=1}^{m}n^{-6/7})$
(recall that $||g_{n}||_{2}\leq 1$ and so $||g_{n}||_{1}\leq \sqrt{2}).$
When $m=\infty $ this sum is infinite, and when $m$ is finite it is $\Omega
(m^{1/7})$. Applying Proposition \ref{p:gap} gives a two-good random
valuation $X$ that satisfies $0<$\textsc{BRev}$(X)<\infty $ (and thus $0<$%
\textsc{DRev}$(X)<\infty $ as well), \textsc{Rev}$(X)=\infty ,$ and \textsc{%
Rev}$_{[m]}(X)=\Omega (m^{1/7})$ for every finite $m.$ For $k>2,$ again, add 
$k-2$ goods with constant valuation $0.$ This proves the two results (for
Theorem \ref{th:infinite gap}(iii) just rescale $X$ to make \textsc{DRev}
equal to $1;$ and the upper bound in Theorem \ref{th:rev-m} is by
Proposition \ref{p:revm}).
\end{proof}

\section{Additive Menu Size\label{s:additive menu}}

We start by proving Theorem \ref{th:srev}, which says that \textsc{MoB} of
selling separately $k$ goods equals the number of goods $k$.

\bigskip

\begin{proof}[Proof of Theorem \protect\ref{th:srev}]
For each good $i$ we have $X_{i}\leq \sum_{\ell }X_{\ell },$ which implies
that\footnote{\label{ft:monotonicity}Use the monotonicity of the one-good
revenue (Hart and Reny 2015 or Hart and Nisan 2017), or Myerson's (1981)
characterization (\ref{eq:one good}).} \textsc{Rev}$(X_{i})\leq \allowbreak $%
\textsc{Rev}$(\sum_{\ell }X_{\ell })=$\textsc{BRev}$(X).$ Summing over $i$
yields \textsc{SRev}$(X)\leq k\cdot $\textsc{BRev}$(X).$

Example 27 in Hart and Nisan (2017) shows that this bound is tight for every 
$k$, even for independent goods.
\end{proof}

\bigskip

Now optimal separate mechanisms sell each good $i$ at a price $p_{i},$ and
so have a menu size of at most $2^{k}-1$ (the buyer can buy any set of goods 
$I\subseteq \{1,\ldots ,k\}$ for the price $\sum_{i\in I}p_{i}$), and yet
Theorem \ref{th:srev} shows that the separate revenue is at most $k$ times
the bundling revenue, rather than $2^{k}-1$ times that (as is the case for
menu size $2^{k}-1,$ and in particular for deterministic mechanisms; see
Theorem \ref{th:drev}). Intuitively, this seems related to the fact that
separate-selling mechanisms have only $k$ \textquotedblleft degrees of
freedom\textquotedblright\ or \textquotedblleft
parameters\textquotedblright\ (the $k$ prices).\footnote{%
This is related to the fact that menu size is defined using the
\textquotedblleft normal" form of a mechanism---its menu---rather than its
other, possibly simpler, descriptions.} To formalize this we introduce a
more refined \textquotedblleft additive menu size\textquotedblright\
complexity measure, as follows.

Let $\mu $ be a $k$-good mechanism with menu $M\subseteq \lbrack
0,1]^{k}\times \mathbb{R}_{+}.$ An \emph{additive representation of }$M$ is
a subset $M_{0}=\{(g_{1},t_{1}),(g_{2},t_{2}),...,(g_{m},t_{m})\}\subseteq M$
of menu entries, which we will refer to as \emph{basic menu entries}, such
that every menu entry $(g,t)$ in $M$ can be represented as a sum of basic
menu entries in $M_{0},$ i.e., $(g,t)=\sum_{n\in N}(g_{n},t_{n})$ for some $%
N\subseteq M_{0},$ and moreover every partial sum $\sum_{n\in N^{\prime
}}(g_{n},t_{n})$ with $N^{\prime }\subset N$ is also a menu entry in%
\footnote{%
Our definition is just one of several possible definitions. Indeed, basic
entries may be combined in other ways---such as taking the allocation
probabilites to be independent (as in Briest et al. 2015), or adding them
and then capping the sum at $1.$ What matters (see the proof of Proposition %
\ref{p:m-star} below) is that any chosen basic entry should yield a
nonnegative payoff (i.e., if $(g,t)$ is part of the set chosen by type $x$
then $g\cdot x-t\geq 0$); the variants mentioned above satisfy this.} $M.$
The \emph{additive menu size} of a mechanism $\mu $ is defined as the
minimal size $|M_{0}|$ of an additive representation of its menu $M.$ Since
taking $M_{0}$ equal to $M$ trivially yields an additive representation, the
additive menu size can thus only be lower than its menu size. For separate
selling of $k$ goods, the additive menu size is at most $k,$ rather than $%
2^{k}-1$: the basic menu entries consist of selling each good by itself at
its price.\footnote{%
More precisely, it is the number of goods whose price is positive.}

The corresponding revenue is

\begin{itemize}
\item \textsc{Rev}$_{[m]\ast }(X),$ the \textquotedblleft \emph{%
additive-menu-size}-$m$" \emph{revenue}, is the maximal revenue that can be
obtained by mechanisms whose additive menu size is at most $m$.
\end{itemize}

Interestingly, the basic properties of the menu size, namely, that menu size 
$1$ yields the bundling revenue, and that the increase in revenue is at most
linear in the menu size (Proposition \ref{p:revm} in Section \ref{sus:menu
size}), hold for the additive menu size as well.

\begin{proposition}
\label{p:m-star}For every $k\geq 2$ and every $k$-good random valuation $X,$

(i) \textsc{Rev}$_{[1]\ast }(X)=$\textsc{Rev}$_{[1]}(X)=$\textsc{BRev}$(X),$
and

(ii) \textsc{Rev}$_{[m]}(X)\leq $\textsc{Rev}$_{[m]\ast }(X)\leq m\cdot $%
\textsc{BRev}$(X)$ for every $m\geq 1.$
\end{proposition}

\begin{proof}
The only claim that is not immediate is the last inequality. Let $M_{0}$
with $|M_{0}|=m$ be a minimal additive representation of the menu. Let $%
(g,t)\in M_{0}$ be a basic menu entry. If the buyer with valuation $x$
chooses $(g,t)$ (i.e., $(g,t)$ is part of the chosen subset $N\subseteq M_{0}
$), then $g\cdot x-t\geq 0$ (otherwise, dropping it from the chosen
subset---i.e., switching to $N\backslash \{(g,t)\},$ which yields an
available menu entry---would strictly increase the buyer's payoff at $x);$
hence $\sum_{i}x_{i}\geq g\cdot x\geq t$ (the first inequality is due to $%
(1,...,1)\geq g$ and $x\geq 0).$ Therefore the total probability\footnote{%
The sum of these probabilities over all basic menu entries may be as high as 
$m,$ as these events need not be disjoint (in contrast to standard menu
items, where they are disjoint).} $\pi $ that $(g,t)$ is chosen is at most $%
\mathbb{P}\left[ \sum_{i=1}^{k}X_{i}\geq t\right] ,$ and so that part of the
expected revenue that comes from $(g,t),$ namely $t\cdot \pi ,$ is at most $%
t\cdot \mathbb{P}\left[ \sum_{i=1}^{k}X_{i}\geq t\right] \leq $\textsc{BRev}$%
(X).$ This holds for each one of the $m$ basic menu entries in $M_{0}.$
\end{proof}

\bigskip

Proposition \ref{p:m-star} thus implies that the results in this paper hold
also for this more refined complexity measure; specifically, in each one of
Theorems \ref{th:infinite gap}--\ref{th:drev} one may replace \textsc{menu
size} with \textsc{menu size}$\ast $. Moreover, by Theorem \ref{th:srev},
this measure captures well the complexity of selling the goods separately:
its \emph{additive} menu size is at most $k.$

\section{Bounded Valuations\label{s:bounded}}

In this section we deal with valuations in bounded domains, i.e., $[L,H]^{k}$
for $0<L<H<\infty .$ Since rescaling valuations by a constant factor of $1/L$
changes the range from $[L,H]^{k}$ to $[1,H/L]^{k}$ without affecting ratios
of revenues, we take without loss of generality $L=1$ and the range $%
[1,H]^{k}$. We first prove Theorem \ref{th:bounded}: for two goods with
valuations in $[1,H]^{2},$ mechanisms need not have more than a \emph{%
polylogarithmic}-in-$H$ menu size in order to obtain arbitrarily good
approximations. It is a direct corollary of the following lemma that shows
how to incur, with an appropriate bounded menu size, only a small loss of
payment for every valuation $x.$

\begin{lemma}
\label{l:logH}Let $k=2.$ For every $H>1$ and $\varepsilon >0$ there exists $%
m=\mathrm{O}(\varepsilon ^{-5}\log ^{2}H)$ such that for every two-good
mechanism $\mu =(q,s)$ whose nonzero payments lie in the range $[1,H]$
(i.e., for each $x$ either $s(x)=0$ or $s(x)\in \lbrack 1,H])$ there exists
a mechanism $\tilde{\mu}=(\tilde{q},\tilde{s})$ with menu size at most $m$
that satisfies $\tilde{s}(x)\geq (1-\varepsilon )s(x)$ for all $x.$
\end{lemma}

\begin{proof}
We will discretize the menu of the given mechanism $\mu .$ Our first step
will be to discretize the payments $s$, and the second to discretize the
allocations $q=(q_{1},q_{2})$.

We start by splitting the range $[1,H]$ into $K$ subranges, each with a
ratio of at most $H^{1/K}$ between its endpoints, where $K$ is chosen so
that $H^{1/K}\leq \varepsilon ^{2}$, i.e., $K=\mathrm{O}(\varepsilon
^{-2}\log H)$. We define a real function $\phi (s)$ by rounding $s$ up to
the top of its range and then multiplying by $1-\varepsilon $. Hence we have 
$(1-\varepsilon )s<\phi (s)<(1-\varepsilon )(1+\varepsilon ^{2})s$. Then for
any $s^{\prime }<s(1-\varepsilon )$ we have $\phi (s)-\phi (s^{\prime
})<(1-\varepsilon )(1+\varepsilon ^{2})s-(1-\varepsilon )s^{\prime
}<s-s^{\prime }$.

Now we take every menu entry $(q,s)$ of the original mechanism and replace $%
s $ with $\phi (s)$. The previous property of $\phi $ ensures that any buyer
who previously preferred $(q,s)$ to some other menu entry $(q^{\prime
},s^{\prime })$ with $s^{\prime }<(1-\varepsilon )s$ still prefers $(q,\phi
(s))$ in the new menu; thus in the new menu he pays $\phi (s^{\prime })$ for
some $s^{\prime }\geq (1-\varepsilon )s$, and $\phi (s^{\prime
})>(1-\varepsilon )s^{\prime }\geq (1-\varepsilon )^{2}s$; his payment in
the new menu is therefore at least $(1-\varepsilon )^{2}$ times his payment
in the original menu.

We now have a menu with only $K$ distinct price levels $s^{1}<\cdots <s^{K}$%
. Before we continue, we scale it down by a factor of $(1-\varepsilon )$,
i.e., multiply both the $q$'s and the $s$'s by $(1-\varepsilon )$. This does
not change the menu choice of any buyer, reduces the payments by a factor of
exactly $1-\varepsilon $, and ensures that $q_{1},q_{2}\leq 1-\varepsilon $.
We now round down each $q_{1}$ and each $q_{2}$ to an integer multiple of $%
\varepsilon /K$, and then add $\varepsilon j/K$ to each menu entry whose
price is $s^{j}$. Notice that rounding down reduces each $q$ by at most $%
\varepsilon /K$, and since higher-paying menu entries got a boost that is at
least $\varepsilon /K$ greater than any lower-paying menu entry, any buyer
that previously chose an entry that pays $s$ can now choose only an entry
that pays some $s^{\prime }\geq s$.

All in all, we have obtained a new mechanism whose payment is at least $%
(1-\varepsilon )^{3}\geq 1-3\varepsilon $ times that of the original one
(and so we redefine the $\varepsilon $ in the proof to be $1/3$ of the $%
\varepsilon $ in the statement). There are $K=\mathrm{O}(\varepsilon
^{-2}\log H)$ price levels and $\varepsilon ^{-1}K=\mathrm{O}(\varepsilon
^{-3}\log H)$ different allocation levels for both $q_{1}$ and $q_{2}$.
However, notice that for a fixed price level $s$ and a fixed $q_{1}$ there
can only be a single value of $q_{2}$ that is actually used in the menu (as
lower ones will be dominated), and so the total number of possible
allocations is $\mathrm{O}(\varepsilon ^{-5}\log ^{2}H)$.
\end{proof}

\bigskip

\begin{proof}[Proof of Theorem \protect\ref{th:bounded}]
Let $X$ be a two-good random valuation with values in $[1,H]^{2},$ and let $%
\mu =(q,s)$ be a two-good mechanism. We have $s(x)\leq q(x)\cdot x\leq 2H$
for every $x\in \lbrack 1,H]^{2};$ and, because the revenue from $X$ is at
least $2$ (obtained, for instance, by selling each good at price $1),$ we
can assume without loss of generality that $R(\mu ;X)\geq 2.$ First, we
eliminate from the menu of $\mu $ all entries whose payment is less than%
\footnote{%
Formally, for every $x$ with $s(x)<2\varepsilon $ we take $(q^{\prime
}(x),s^{\prime }(x))$ to be a maximizer of $q(y)\cdot x-s(y)$ over all $y$
such that either $s(y)=0$ or $s(y)\geq 2\varepsilon .$} $2\varepsilon ;$ any
type $x$ with $s(x)<2\varepsilon $ then either pays $0,$ or some $s(y)\geq
2\varepsilon .$ The loss in revenue, if any, is thus at most $2\varepsilon
\cdot \mathbb{P}\left[ s(X)<2\varepsilon \right] \leq 2\varepsilon .$ Let $%
\mu ^{\prime }=(q^{\prime },s^{\prime })$ denote the resulting mechanism;
then the range of its nonzero payments is $[2\varepsilon ,2H].$ Applying
Lemma \ref{l:logH} to $\mu ^{\prime }$ yields a new mechanism $\tilde{\mu}=(%
\tilde{q},\tilde{s})$ with a menu of size \textrm{O}$(\varepsilon ^{-5}\log
^{2}(H/\varepsilon )),$ such that $\tilde{s}(x)\geq (1-\varepsilon )s(x)$
for all $x,$ and thus 
\begin{eqnarray*}
R(\tilde{\mu};X) &\geq &(1-\varepsilon )R(\mu ^{\prime };X)\geq
(1-\varepsilon )(R(\mu ;X)-2\varepsilon ) \\
&\geq &(1-2\varepsilon )R(\mu ;X)
\end{eqnarray*}%
(recall that $R(\mu ;X)\geq 2).$
\end{proof}

\bigskip

Notice that the polylogarithmic dependence of $m$ on $H$ is
\textquotedblleft about right\textquotedblright\ since the valuation $X$
induced by the first $m$ points in the construction of Proposition \ref%
{qn-seq} (used for proving Theorem \ref{th:infinite gap}(ii)) has $H=m^{%
\mathrm{O}(m)}$, and the $\Omega (m^{1/7})$ gap between \textsc{Rev}$(X)$
and \textsc{Rev}$_{[1]}(X)$ implies that for, say, $m=\mathrm{O}((\log
H)^{1/8}),$ we get \textsc{Rev}$_{[m]}(X)=\mathrm{o}($\textsc{Rev}$(X))$.

For more than two goods, i.e., $k>2,$ we obtain the somewhat weaker result
that the menu size need only be polynomial in $H$.

\begin{proposition}
\label{th:additive-delta} For every $k\geq 2$ and $\varepsilon >0$ there is $%
m_{0}=(k/\varepsilon )^{\mathrm{O}(k)}$ such that for every $k$-good random
valuation $X$ with values in $[0,1]^{k}$ and every $m\geq m_{0}$, 
\[
\text{\textsc{Rev}}_{[m]}(X)\geq \text{\textsc{Rev}}(X)-\varepsilon . 
\]
\end{proposition}

This result is directly implied by the following lemma.

\begin{lemma}
Let $m=(n+1)^{k}-1,$ where $n\geq 1$ is an integer. Then for every $k$-good
random valuation $X$ with values in $[0,1]^{k}$, 
\[
\text{\textsc{Rev}}_{[m]}(X)\geq \text{\textsc{Rev}}(X)-\frac{2k}{\sqrt{n}}.
\]
\end{lemma}

\begin{proof}
Let $X$ have values in $[0,1]^{k},$ and let $\mu =(q,s)$ be a mechanism.

Define a new mechanism $\tilde{\mu}=(\tilde{q},\tilde{s})$ as follows: for
each $x\in \lbrack 0,1]^{n},$ let $\tilde{q}(x)$ be the rounding up of $q(x)$
to the $1/n$-grid on $[0,1]^{k}$, and let $\tilde{s}(x):=(1-1/\sqrt{n})s(x).$
Since $\tilde{q}$ can take at most $(n+1)^{k}$ different values, the menu
size of $\tilde{\mu}$ is at most $(n+1)^{k}-1=m$.

If $\tilde{q}(x)\cdot x-\tilde{s}(x)\leq \tilde{q}(y)\cdot x-\tilde{s}(y),$
then (recall that $q(x)\cdot x-s(x)\geq q(y)\cdot x-s(y)$) we must have $%
(1/n)\sum_{i=1}^{k}x_{i}\geq (1/\sqrt{n})(s(x)-s(y));$ hence $s(y)\geq
s(x)-k/\sqrt{n}$ (since $\sum_{i}x_{i}\leq k),$ which implies that the
seller's revenue at $x$ from $\tilde{\mu}$ must be $\geq (1-1/\sqrt{n}%
)(s(x)-k/\sqrt{n}).$ Therefore $R(\tilde{\mu};X\mathcal{)}\geq (1-1/\sqrt{n}%
)R(\mu ;X)-k/\sqrt{n}\geq R(\mu ;X)-2k/\sqrt{n}$ (since $R(\mu ;X)\leq
\sum_{i}x_{i}\leq k).$
\end{proof}

\bigskip

From Proposition \ref{th:additive-delta} we can derive an essentially
equivalent multiplicative approximation result.

\begin{proposition}
\label{th:1-delta} For every $k\geq 2$, $\varepsilon >0$, and $H>1,$ there
is $m_{0}=(H/\varepsilon )^{\mathrm{O}(k)}$ such that for every $k$-good
random valuation $X$ with values in $[1,H]^{k}$ and every $m\geq m_{0}$, 
\[
\text{\textsc{Rev}}_{[m]}(X)\geq (1-\varepsilon )\cdot \text{\textsc{Rev}}%
(X). 
\]
\end{proposition}

\begin{proof}
We first rescale $[1,H]$ to $[1/H,1]$, which for multiplicative
approximations is the same. We then design a mechanism that gives an
additive approximation to within $\varepsilon k/H$, which, by Proposition %
\ref{th:additive-delta}, requires a menu size $m$ as stated. Now, since each 
$X_{i}$ is bounded from below by $1/H$, the revenue of $X$ is at least $k/H$
(each good is sold for sure at the price $1/H),$ and thus an $\varepsilon k/H
$-additive approximation is also a $(1-\varepsilon )$-multiplicative
approximation, as required.
\end{proof}

%TCIMACRO{\TeXButton{appendix}{\appendix}}%
%BeginExpansion
\appendix%
%EndExpansion

\section{Appendix}

\subsection{Two-Good Deterministic Mechanisms\label{ap:beta}}

Using the formula of Theorem \ref{th:beta} we can show that the Multiple of
Basic revenue for two-good deterministic mechanisms equals precisely the $%
d_{2}=5/2$ bound of Proposition \ref{p:drev/brev} (see Remark \ref{r:d_k}%
(c)).

\begin{proposition}
\label{p:5/2}For $k=2$ goods, 
\[
\text{\textsc{MoB}}(\text{\textsc{deterministic}};\text{ }2\text{ goods})=%
\frac{5}{2}.
\]
\end{proposition}

\begin{proof}
We compute the supremum of \textsc{MoB}$(\mathcal{\mu }),$ as given by
Theorem \ref{th:beta}, over all deterministic mechanisms $\mu .$ Such a
mechanism is given by nonnegative prices $p_{1},p_{2},$ and $p_{12}$ for
good $1,$ good $2,$ and the bundle, respectively (thus $b(x)=\max
\{0,x_{1}-p_{1},x_{2}-p_{2},x_{1}+x_{2}-p_{12}\})$. Without loss of
generality we assume that $p_{1}\leq p_{2}\leq p_{12};$ the first inequality
because we can interchange the two coordinates, and the second because if $%
p_{i}>p_{12}$ then the menu entry $x_{i}-p_{i}$ is never chosen, and so
replacing $p_{i}$ with $p_{i}^{\prime }:=p_{12}$ does not affect the
revenue. We have four cases:

\begin{itemize}
\item If $p_{1}>0$ then \textsc{MoB}$(\mu
)=(p_{1}-0)/v(p_{1})+(p_{2}-p_{1})/v(p_{2})+(p_{12}-p_{2})/v(p_{12}).$ Now $%
v(p_{1})=p_{1}$ (attained at $x=(p_{1},0)$) and $v(p_{2})=p_{2}$ (attained
at $x=(0,p_{2})$)$;$ as for $v(p_{12}),$ if $s(x)=p_{12}$ then $%
x_{1}+x_{2}-p_{12}\geq x_{i}-p_{i}$ for $i=1,2,$ which implies $x_{3-i}\geq
p_{12}-p_{i}\geq p_{12}-p_{2},$ and so $x_{1}+x_{2}\geq 2(p_{12}-p_{2}).$
Therefore \textsc{MoB}$(\mu )\leq 1+1+1/2=5/2.$

\item If $p_{1}=0<p_{2}$ then \textsc{MoB}$(\mu
)=(p_{2}-0)/v(p_{2})+(p_{12}-p_{2})/v(p_{12})\leq 1+1/2=3/2.$

\item If $p_{1}=p_{2}=0<p_{12}$ then \textsc{MoB}$(\mu )\leq 1.$

\item If $p_{1}=p_{2}=p_{12}=0$ then \textsc{MoB}$(\mu )=0.$
\end{itemize}

Thus \textsc{MoB}$(\mu )\leq 5/2$ in all cases; taking, say, $p_{1}=1,$ $%
p_{2}=H$, and $p_{12}=H^{2}$ for large\footnote{%
Alternatively, use the bound of Proposition \ref{p:drev/brev} (see Remark %
\ref{r:d_k}(b)).} $H$ shows that $\sup_{\mu }\text{\textsc{MoB}}(\mu )$ over
all deterministic mechanisms $\mu $ is indeed $5/2.$
\end{proof}

\bigskip

For \emph{separate-selling} mechanisms we have in addition $%
p_{12}=p_{1}+p_{2},$ and then $v(p_{1}+p_{2})=p_{1}+p_{2}$ (attained at $%
x=(p_{1},p_{2})),$ and so \textsc{MoB}$(\mu
)=1+1-p_{1}/p_{2}+p_{1}/(p_{1}+p_{2}),$ which is less than $2,$ but can be
made arbitrarily close to $2$ by taking, say, $p_{1}=1$ and $p_{2}=H$ for
large $H.$ This shows that \textsc{MoB}$($\textsc{separate}$;$ $2$ goods$)=2;
$ cf. Theorem \ref{th:srev}. For \emph{symmetric} deterministic mechanisms
we have $p_{1}=p_{2},$ and so \textsc{MoB}$(\mu )\leq
p_{1}/p_{1}+(p_{12}-p_{1})/(2(p_{12}-p_{2}))=3/2,$ with equality for, say, $%
p_{1}=p_{2}=1$ and $p_{12}=2$ (which is in fact a symmetric separate-selling
mechanism). Thus \textsc{MoB}$($\textsc{symmetric deterministic}$;$ $2$ goods%
$)=\text{\textsc{MoB}}(\text{\textsc{symmetric separate}}$; $2$ goods$)=3/2.$

\subsection{The Multiple of Separate Revenue (\textsc{MoS})\label{ap:gamma}}

Our \textsc{MoB} measure takes as basic revenue the bundling revenue,
obtained by menu-size-$1.$ We now consider using the separate revenue
instead:%
\[
\text{\textsc{MoS}}(\mathcal{N};\mathbb{X}):=\sup_{x\in \mathbb{X}}\frac{%
\mathcal{N}\text{-\textsc{Rev}}(X)}{\text{\textsc{SRev}}(X)}
\]%
(\textsc{MoS} stands for \textquotedblleft Multiple of Separate revenue").

We start with a simple comparison between the bundling and separate revenues.

\begin{proposition}
\label{p:b/s<=k}For every $k\geq 2,$ 
\[
\text{\textsc{MoS}}(\text{\textsc{bundled}};\text{ }k\text{ goods})\leq k.
\]
\end{proposition}

\begin{proof}
Let \textsc{BRev}$(X)$ be achieved for a bundle price of $p$. If the
separate auction offers each good at a price of $p/k$ then whenever $%
\sum_{i}x_{i}\geq p$ we have $x_{i}\geq p/k$ for some $i$, and so one of the 
$k$ goods will be acquired in the separate auction; thus \textsc{BRev}$%
(X)\leq k\cdot $\textsc{SRev}$(X).$
\end{proof}

\bigskip

This is tight for $k=2.$

\begin{example}
\label{ex:b/s=k}Let $X_{1}$ be distributed uniformly on $[0,1],$ and
consider the two-good random valuation $X=(X_{1},1-X_{1}).$ The bundling
revenue is $1$, since the bundle is always worth $1$ to the buyer. Each good
is distributed uniformly on $[0,1]$ and so the optimal revenue from each
good is, by (\ref{eq:one good}), $1/4$ (obtained at price $1/2$).
\end{example}

\bigskip

For larger values of $k$, we can get a stronger result.

\begin{proposition}
\label{p:b/s<=logk}There exists a constant $c<\infty $ such that for every $%
k\geq 2$ and every $k$-good random valuation $X,$ 
\[
\text{\textsc{MoS}}(\text{\textsc{bundled}};\text{ }k\text{ goods})\leq
c\log k.
\]
\end{proposition}

\begin{proof}
Let \textsc{BRev}$(X)$ be achieved for bundle price $p$. We first assume
without loss of generality that the support of $X$ contains only points $x$
with $\sum_{i}x_{i}=p$ or $\sum_{i}x_{i}=0$. (This is without loss of
generality, since the random variable $X^{\prime }$ defined by $X^{\prime
}:=0$ when $\sum_{i}X_{i}<p$ and $X:=(p/\sum_{i}X_{i})X$ satisfies \textsc{%
BRev}$(X^{\prime })=$\textsc{BRev}$(X),$ while \textsc{SRev}$(X^{\prime
})\leq $\textsc{SRev}$(X)$ because $X^{\prime }\leq X$ everywhere.\footnote{%
See footnote \ref{ft:monotonicity} above.}) We now make another assumption
without loss of generality, namely, that $\sum_{i}X_{i}=p$ (and so \textsc{%
BRev}$(X)=p$). (This is without loss of generality because if we replace $X$
with its conditional on $\sum_{i}x_{i}=p,$ then all revenues are just
rescaled by a factor of $1/\mathbb{P}\left[ \sum_{i}X_{i}=p\right] $.)

At this point there are two different ways to proceed; we present both, as
they may lead to different extensions.

\emph{Proof 1:} Let $e_{i}:=\mathbb{E}[X_{i}]$ be the expected value of good 
$i$; then (using our assumptions) $\sum_{i}e_{i}=p$. The claim is that good $%
i$ can be sold in a separate auction yielding a revenue of at least $%
(e_{i}-p/(2k))/(2(1+\log _{2}k))$. The result is then implied by summing
over all $i$.

Indeed, split the range of values of $X_{i}$ into $(2+\log _{2}k)$
subranges: a \textquotedblleft low\textquotedblright\ subrange for which $%
X_{i}\leq p/(2k),$ and, for each $j=0,\ldots ,\log _{2}k,$ a subrange where $%
p/(2^{j+1})<X_{i}\leq p/(2^{j})$ (notice that since $X_{i}\leq p$ we have
covered the whole support of $X_{i}$). The low subrange contributes at most $%
p/(2k)$ to the expectation of $X_{i}$, and thus one of the other $1+\log
_{2}k$ subranges contributes at least $((e_{i}-p/(2k))/(1+\log _{2}k)$ to
this expectation. The lower bound of this subrange, $p/(2^{j+1})$, is
smaller by a factor of at most $2$ than any value in the subrange, and so
setting it as the price for good $i$ yields a revenue that is at least half
of the contribution of this subrange to the expectation, i.e., at least $%
((e_{i}-p/(2k))/(2(1+\log _{2}k))$.

\emph{Proof 2:} Let $r_{i}:=$\textsc{Rev}$(X_{i})=\sup_{t>0}t\cdot
(1-F_{i}(t))$ (where $F_{i}$ denotes the cumulative distribution function of 
$X_{i});$ then $1-F_{i}(t)\leq r_{i}/t$ and so (recall that $X_{i}\leq p$
because $\sum_{i}X_{i}=p)$%
\[
\mathbb{E}\left[ X_{i}\right] =\int_{0}^{\infty }(1-F_{i}(t))\mathrm{d}t\leq
\int_{0}^{r_{i}}1\mathrm{d}t+\int_{r_{i}}^{p}\frac{r_{i}}{t}\mathrm{d}%
t=r_{i}(1+\ln p-\ln r_{i}).
\]%
Averaging over $i$ and using the concavity in $r$ of the function $r(1+\ln
p-\ln r)$ yields 
\[
\frac{p}{k}=\frac{1}{k}\sum_{i=1}^{k}\mathbb{E}\left[ X_{i}\right] \leq 
\frac{s}{k}\left( 1+\ln p-\ln \frac{s}{k}\right) ,
\]%
where $s:=\sum_{i}r_{i}$. Thus $p/s\leq 1+\ln (p/s)+\ln k$, from which it
follows that\footnote{%
The function $x-\ln x-1-\ln k$ is increasing in $x,$ and is positive at $%
x=4\ln k$ (because $k\geq 2$ implies $k^{3}/\ln k>4e).$} \textsc{BRev}$(X)/$%
\textsc{SRev}$(X)=p/s<4\ln k.$
\end{proof}

\begin{corollary}
\label{c:drev/srev<=2^k*logk}There exists a constant $c<\infty $ such that
for every $k\geq 2,$%
\[
\text{\textsc{MoS}}(\text{\textsc{deterministic}};\text{ }k\text{ goods}%
)\leq c2^{k}\log k.
\]
\end{corollary}

For the special case of $k=2$ goods, we have a somewhat tighter bound.

\begin{proposition}
\label{p:d/s<=3} Let $k=2.$ Then 
\[
\text{\textsc{MoS}}(\text{\textsc{deterministic}};\text{ }2\text{ goods}%
)\leq 3.
\]
\end{proposition}

\begin{proof}
A deterministic mechanism has at most three menu entries: either selling
just one of the goods, or selling the bundle. The portion of the revenue
that comes from those types that buy only good $i$ cannot exceed \textsc{Rev}%
$(X_{i}),$ and the portion that comes from those that buy the bundle cannot
exceed \textsc{BRev}$(X);$ in total, \textsc{DRev}$(X)\leq $\textsc{SRev}$%
(X)+$\textsc{BRev}$(X)$. The proof is completed using Proposition \ref%
{p:b/s<=k}.
\end{proof}

\bigskip

We now study \textsc{MoS}; the analysis is analogous to the one carried out
with respect to the bundling revenue in Sections \ref{s:beta}--\ref{s:gap},
but we now use the maximum norm $||x||_{\infty }=\max_{i}|x_{i}|$ instead of
the $1$-norm.

We have

\begin{theorem}
\label{th:gamma}Let $\mu =(q,s)$ be a $k$-good mechanism. Then%
\[
\frac{1}{k}\int_{0}^{\infty }\frac{1}{w(t)}~\mathrm{d}t\leq \text{\textsc{MoS%
}}(\mu )\leq \int_{0}^{\infty }\frac{1}{w(t)}~\mathrm{d}t, 
\]
where for every $t>0$ we define%
\[
w(t)%
%TCIMACRO{\TeXButton{:=}{{\;:=\;}}}%
%BeginExpansion
{\;:=\;}%
%EndExpansion
\inf \{||x||_{\infty }:x\in \mathbb{R}_{+}^{k}\text{ and }s(x)\geq t\}. 
\]
\end{theorem}

Unlike Theorem \ref{th:beta}, here we do not get a sharp formula for \textsc{%
MoS}, but only an integral that is within a factor of $k$ from it (see
Remark \ref{r:mos=int(1/w)}(b) below).

\begin{proof}
Let $\gamma :=\int_{0}^{\infty }1/w(t)~\mathrm{d}t.$

First, for every $t>0$ we have 
\begin{eqnarray*}
\mathbb{P}\left[ s(X)\geq t\right] &\leq &\mathbb{P}\left[ ||X||_{\infty
}\geq w(t)\right] =\mathbb{P}\left[ \cup _{i}\{X_{i}\geq w(t)\}\right] \leq
\sum_{i}\mathbb{P}\left[ X_{i}\geq w(t)\right] \\
&\leq &\sum_{i}\frac{\text{\textsc{Rev}}(X_{i})}{w(t)}=\frac{\text{\textsc{%
SRev}}(X)}{w(t)}.
\end{eqnarray*}%
Integrating over $t$ yields $R(\mu ,X)\leq \gamma \cdot $\textsc{SRev}$(X),$
proving that \textsc{MoS}$(\mu )\leq \gamma .$

Second, we show that for every $\gamma ^{\prime }<\gamma $\ there exists a $k
$-good random valuation $X$ such that $0<$\textsc{SRev}$(X)<\infty $\ and $%
R(\mu ;X)/\text{\textsc{SRev}}(X)>\gamma ^{\prime }/k.$ Let $%
0=t_{0}<t_{1}<...<t_{N}<t_{N+1}=\infty $ with $%
0=w(t_{0})<w(t_{1})<w(t_{2})<...<w(t_{N})<w(t_{N+1})=\infty $ be such that%
\[
\gamma ^{\prime \prime }:=\sum_{n=1}^{N}\frac{t_{n}-t_{n-1}}{w(t_{n})}%
>\gamma ^{\prime }.
\]%
Let $\varepsilon >0$ be small enough so that $\gamma ^{\prime \prime
}>(1+\varepsilon )\gamma ^{\prime }$ and $w(t_{n+1})>(1+\varepsilon )w(t_{n})
$ for all $1\leq n\leq N,$ and choose for each $1\leq n\leq N$ a point $%
x_{n}\in \mathbb{R}_{+}^{k}$ such that $s(x_{n})\geq t_{n}$ and $%
w(t_{n})\leq ||x_{n}||_{\infty }<(1+\varepsilon )w(t_{n})$; then%
\begin{equation}
\sum_{n=1}^{N}\frac{t_{n}-t_{n-1}}{||x_{n}||_{1}}>\sum_{n=1}^{N}\frac{%
t_{n}-t_{n-1}}{w(t_{n})(1+\varepsilon )}=\frac{\gamma ^{\prime \prime }}{%
1+\varepsilon }>\gamma ^{\prime }.  \label{eq:sum-gamma'}
\end{equation}

Let $X$ be a random variable with support $\{x_{1},...,x_{N}\}$ and
distribution $\mathbb{P}\left[ X=x_{n}\right] =\xi _{1}/\xi _{n}-\xi
_{1}/\xi _{n+1}$ for every $1\leq n\leq N,$ where $\xi
_{n}:=||x_{n}||_{\infty }$ and we put $\xi _{N+1}:=\infty ;$ thus $\mathbb{P}%
\left[ X\in \{x_{n},...,x_{N}\}\right] =\xi _{1}/\xi _{n}$ for every $n\geq
1.$

Consider good $i.$ For every $u\in (\xi _{n-1},\xi _{n}]$ (with $1\leq n\leq
N)$ we have 
\[
u\cdot \mathbb{P}\left[ X_{i}\geq u\right] \leq u\cdot \mathbb{P}\left[ X\in
\{x_{n},...,x_{N}\}\right] =u\frac{\xi _{1}}{\xi _{n}}\leq \xi _{1} 
\]%
(because $X=x_{j}$ for some $j\leq n-1$ implies $X_{i}\leq ||x_{j}||_{\infty
}\leq ||x_{n-1}||_{\infty }=\xi _{n-1}<u).$ Therefore \textsc{Rev}$%
(X_{i})=\sup_{u>0}u\cdot \mathbb{P}\left[ X_{i}\geq u\right] \leq \xi _{1}$\
for every good $i,$ and so \textsc{SRev}$(X)\leq k\xi _{1}$ (which is
finite; also \textsc{SRev}$(X)>0$ because $X$ does not vanish).

Finally, the revenue of $R(\mu ;X)$ that $\mu $ gets from $X$ is%
\begin{eqnarray*}
R(\mu ;X) &\geq &\sum_{n=1}^{N}s(x_{n})\mathbb{P}\left[ X=x_{n}\right] \geq
\sum_{n=1}^{N}t_{n}\left( \frac{\xi _{1}}{\xi _{n}}-\frac{\xi _{1}}{\xi
_{n+1}}\right) \\
&=&\xi _{1}\sum_{n=1}^{N}\frac{t_{n}-t_{n-1}}{\xi _{n}}>\xi _{1}\gamma
^{\prime }=\frac{\gamma ^{\prime }}{k}\cdot k\xi _{1}\geq \frac{\gamma
^{\prime }}{k}\cdot \text{\textsc{SRev}}(X)
\end{eqnarray*}%
(recall (\ref{eq:sum-gamma'})).
\end{proof}

\begin{remark}
\label{r:mos=int(1/w)}\emph{(a) }As in Theorem \ref{th:beta} (see Remark \ref%
{r:mob=int(1/v)} following its proof), the random valuation $X$ in the
second part of the proof may be taken so that its values are in $[0,1]^{k}$
and its support is at most the size of the menu of $\mu $.

\emph{(b) }The gap of $k$ in Theorem \ref{th:gamma} is correct. Take two
goods. For $\mu $ that sells the bundle at the price of $1$ we have $w(1)=1/2
$ (attained at $x=(1/2,1/2))$ and so $\gamma (\mu )=2;$ the two-good random
valuation $X$ of Example \ref{ex:b/s=k} has $R(\mu ;X)/$\textsc{SRev}$%
(X)=1/(1/2)=\gamma (\mu )$. For $\mu $ that sells each good separately for
the price of $1/2$ we have $w(1/2)=1/2$ and $w(1)=1,$ and so\footnote{%
It is easy to see that $\gamma (\mu )=k$ for every $k$-good mechanism $\mu $
that sells the goods separately at positive prices.} $\gamma (\mu )=2$, but $%
R(\mu ;X)/$\textsc{SRev}$(X)\leq 1=\gamma (\mu )/k$ for any $X$ (with
equality for, say, the constant valuation $(1/2,1/2)).$

\emph{(c) }Recalling the definition of $v(t)$ in Theorem \ref{th:beta}, we
have $1/v(t)\leq 1/w(t)\leq k/v(t)$ for every $t$ (because $||x||_{1}\geq
||x||_{\infty }\geq ||x||_{1}/k),$ and so for every mechanism $\mu $ we have 
\textsc{MoB}$(\mu )\leq \int 1/w(t)~\mathrm{d}t\leq k\cdot \text{\textsc{MoB}%
}(\mu ).$

\emph{(d) }We can take as benchmark the maximum of the two one-dimensional
mechanisms, bundled and separate (cf. Babaioff et al. 2014). Thus, putting%
\[
\text{\textsc{MoBS}}(\mu \mathbb{)}:=\sup_{X}\frac{R(\mu ;X)}{\max \{\text{%
\textsc{BRev}}(X),\text{\textsc{SRev}}(X)\}},
\]%
we have%
\begin{equation}
\frac{1}{k}\int_{0}^{\infty }\frac{1}{w(t)}~\mathrm{d}t\leq \text{\textsc{%
MoBS}}(\mu \mathbb{)}\leq \int_{0}^{\infty }\frac{1}{v(t)}~\mathrm{d}t.
\label{eq:mobs}
\end{equation}%
Indeed, in the second part of the proof of Theorem \ref{th:gamma} above, for
every $u\in (k\xi _{n-1},k\xi _{n}],$ 
\[
u\cdot \mathbb{P}\left[ \sum_{i=1}^{k}X_{i}\geq u\right] \leq u\cdot \mathbb{%
P}\left[ X\in \{x_{n},...,x_{N}\}\right] =u\frac{\xi _{1}}{\xi _{n}}\leq
k\xi _{1}
\]%
(because $X=x_{j}$ for some $j\leq n-1$ implies $\sum_{i}X_{i}\leq
k||x_{j}||_{\infty }\leq k||x_{n-1}||_{\infty }=k\xi _{n-1}<u$), and so 
\textsc{BRev}$(X)=\sup_{u>0}u\cdot \mathbb{P}\left[ X_{i}\geq u\right] \leq
k\xi _{1}$ as well, which yields the first inequality in (\ref{eq:mobs}).
For the second inequality we use \linebreak \textsc{MoBS}$(\mu \mathbb{)}%
\leq $\textsc{MoB}$(\mu \mathbb{)\leq }\int 1/v$ (which, by (c) above,
yields a better inequality than \textsc{MoBS}$(\mu \mathbb{)}\leq $\textsc{%
MoS}$(\mu \mathbb{)\leq }\int 1/w$).
\end{remark}

\bigskip

The analogous result to the construction of Section \ref{s:gap} is

\begin{proposition}
\label{p:gap-infinity}Let $(g_{n})_{n=0}^{N}$ be a finite or countably
infinite sequence in $[0,1]^{k}$ starting with $g_{0}=(0,...,0),$ and let $%
(y_{n})_{n=1}^{N}$ be a sequence of vectors in $\mathbb{R}_{+}^{k}$ such that%
\[
\mathrm{gap}_{n}:=\min_{0\leq j<n}(g_{n}-g_{j})\cdot y_{n}>0
\]%
for all $n\geq 1.$ Then for every $\varepsilon >0$ there exist a sequence $%
(t_{n})_{n=1}^{N}$ of positive real numbers, a $k$-good mechanism $\mu $
with menu $\{(g_{n},t_{n})\}_{n=1}^{N},$ and a $k$-good random valuation $X$
with $0<$\textsc{BRev}$(X)<\infty ,$ such that%
\[
\text{\textsc{MoS}}(X)\geq \text{\textsc{MoBS}}(X)>(1-\varepsilon )\frac{1}{k%
}\sum_{n=1}^{N}\frac{\mathrm{gap}_{n}}{||y_{n}||_{\infty }}.
\]
\end{proposition}

The proof is omitted, as it is identical to that of Proposition \ref{p:gap},
except that it uses throughout the $\infty $-norm instead of the $1$-norm
(and the construction of the appropriate random valuation is as in Theorem %
\ref{th:gamma} and Remark \ref{r:mos=int(1/w)}(d) above instead of Theorem %
\ref{th:beta}).

As a consequence, for deterministic mechanisms we get (see Corollary \ref%
{c:drev/srev<=2^k*logk} for the opposite inequality):

\begin{proposition}
\label{p:drev/brevsrev}For every $k\geq 2,$%
\[
\text{\textsc{MoS}}(\text{\textsc{deterministic}};\text{ }k\text{ goods}%
)\geq \text{\textsc{MoSB}}(\text{\textsc{deterministic}};\text{ }k\text{
goods})\geq \frac{2^{k}-1}{k}.
\]
\end{proposition}

\begin{proof}
We proceed exactly as in the proof of Proposition \ref{p:drev/brev}, but now
we have $||y_{n}||_{\infty }=1,$ and so we get%
\[
\frac{1}{k}\sum_{n=1}^{2^{k}-1}\frac{\mathrm{gap}_{n}}{||y_{n}||_{\infty }}=%
\frac{1}{k}\sum_{\ell =1}^{k}\binom{k}{\ell }=\frac{2^{k}-1}{k}. 
\]
\end{proof}

\bigskip

For $k=2$ goods, the supremum of $\int 1/w$ over all deterministic
mechanisms equals $3$ (attained in the limit as $H\rightarrow \infty $ by
prices $p_{1}=1,p_{2}=H,p_{12}=H^{2};$ cf. the proof of Proposition \ref%
{p:5/2}). Thus, 
\begin{eqnarray*}
\frac{3}{2} &\leq &\text{\textsc{MoS}}(\text{\textsc{deterministic}; }2\text{
\emph{goods}})\leq 3 \\
\frac{3}{2} &\leq &\text{\textsc{MoSB}}(\text{\textsc{deterministic}; }2%
\text{ \emph{goods}})\leq \frac{5}{2}
\end{eqnarray*}%
(cf. Proposition \ref{p:5/2}, which shows that \textsc{MoB} is exactly $5/2$%
).

\subsection{The Unit-Demand Model\label{ap:unit demand}}

In this section we briefly compare our model to the unit-demand model that
is considered in many papers. There are $k$ goods for sale and a single
buyer. There are two basic differences between our model and the unit-demand
one. First, in the unit-demand model, the buyers are modeled as having
unit-demand valuations. Additionally, the unit-demand model requires the
mechanism to offer only single goods, rather than bundles of goods as in our
model. This second restriction does not turn out to matter.

More formally, in the unit demand model there is a single buyer with a unit
demand valuation; i.e., the valuation of a set $I\subseteq \{1,\ldots ,k\}$
of goods is $\max_{i\in I}x_{i}$ (rather than $\sum_{i\in I}x_{i}$). A
deterministic mechanism in this setting would offer a price $p_{i}$ for each
good $i$. For unit-demand buyers this is equivalent to a completely general
deterministic mechanism as there is no need to offer prices for bundles
since the buyer is not interested in them. Thus, for example, a mechanism
asking price $p_{1}$ for good $1$, price $p_{2}$ for good $2$, and price $%
p_{12}$ for both goods would be the same as asking price $\min
\{p_{1},p_{12}\}$ for good $1$ and price $\min \{p_{1},p_{12}\}$ for good $2$%
.

A randomized mechanism in this model is allowed to offer a set of lotteries,
each with its own price, where a lottery is a vector of probabilities $%
\alpha _{1},\ldots ,\alpha _{k}$ of getting the goods, with $\sum_{i}\alpha
_{i}\leq 1$ (in contrast to our additive buyer, where $q_{i}\leq 1$ for each 
$i$). Again, for unit-demand buyers this is equivalent to general randomized
mechanisms that are also allowed to offer lotteries for bundles of goods.
For example, a menu entry offering the lottery \textquotedblleft good $1$
with probability $2/9$; good $2$ with probability $3/9$; and both goods with
probability $4/9$" at a certain price can be replaced by the two menu
entries \textquotedblleft good $1$ with probability $6/9$; good $2$ with
probability $3/9$" and \textquotedblleft good $1$ with probability $2/9$;
good $2$ with probability $7/9$," each at the same price as in the original
menu entry.

Let us use the notation \textsc{Rev}$^{\QTR{sc}{UD}}(X)$ to denote the
revenue obtainable from a unit-demand buyer with a $k$-good random valuation 
$X$. Similarly \textsc{DRev}$^{\QTR{sc}{UD}}(X)$ denotes the revenue
achievable by deterministic mechanisms. We can compare these revenues to
those achievable in our model from an additive buyer whose valuation for the 
$k$ goods is given by the same $X$.

\begin{proposition}
For every $k\geq 2$ and every $k$-good random valuation $X,$

(i) \textsc{Rev}$^{\QTR{sc}{UD}}(X)\leq $\textsc{Rev}$(X)\leq k\cdot $%
\textsc{Rev}$^{\QTR{sc}{UD}}(X)$, and

(ii) \textsc{DRev}$^{\QTR{sc}{UD}}(X)\leq $\textsc{DRev}$(X)\leq k2^{k}\cdot 
$\textsc{DRev}$^{\QTR{sc}{UD}}(X)$.
\end{proposition}

\begin{proof}
The lower bounds in both cases are obtained by noting that any mechanism in
the unit-demand model offers only unit-demand menu entries, and for these
both the unit-demand buyer and the additive buyer have the same preferences;
thus offering the same menu in our setting gives exactly the same revenue as
it does in the unit-demand setting.

For the upper bound for randomized mechanisms in (i), notice that if we
replace each menu entry $((g_{1},...,g_{k});t)$ in our model (where $0\leq
g_{i}\leq 1$ for each $i$) by the menu entry $((g_{1}/k,\ldots ,g_{k}/k);t/k)
$, then we do not change the preferences of the buyer between the different
menu entries, and thus the revenue drops by a factor of exactly $k$.
However, the new mechanism gives only unit-demand allocations (because $%
g_{1}/k+...+g_{k}/k\leq 1),$ and for these the unit-demand buyer and the
additive buyer behave the same.

For the upper bound for deterministic mechanisms in (ii), consider a
deterministic mechanism in our model. Since it has at most $2^{k}-1$ menu
entries, a fraction of at least $2^{-k}$ of the revenue must come from one
of them, which allocates, say, a set $I$ of goods. A mechanism that offers
to sell only this set $I$ of goods at the same price $t$ as the original
mechanism did will thus make at least a $2^{-k}$ fraction of the revenue of
the original one. Now consider the unit-demand mechanism that offers each
one of the goods in $I$ at the price $t/|I|;$ whenever the additive buyer in
the additive mechanism buys $I$ we are guaranteed that his value for at
least one of the goods in $I$ is at least $t/|I|$, in which case the
unit-demand buyer will also acquire that good at $t/|I|$ in the unit-demand
mechanism.
\end{proof}

\bigskip

The interesting gap in the above proposition is the exponential one for
deterministic mechanisms in (ii), and indeed we can show that this is
essentially tight.

\begin{proposition}
For every $k\geq 2,$ 
\[
\sup_{X}\frac{\text{\textsc{DRev}}(X)}{\text{\textsc{DRev}}^{\QTR{sc}{UD}}(X)%
}\geq \frac{2^{k}-1}{k}. 
\]
\end{proposition}

\begin{proof}
For every $X$ we have \textsc{DRev}$^{\QTR{sc}{UD}}(X)\leq $\textsc{SRev}$%
(X) $ because the good prices used in any deterministic mechanism in the
unit-demand model can only yield more revenue in our additive model where
the buyer may buy more than a single good. Use Proposition \ref%
{p:drev/brevsrev} in Appendix \ref{ap:gamma}.
\end{proof}

\bigskip

Despite the exponential separation, for fixed $k$ it is constant, and so a
super-constant separation between randomized and deterministic mechanisms in
our setting is equivalent to the same separation in the unit-demand setting.

\subsection{More Than One Buyer\label{ap:n-buyers}}

This paper has concentrated on a single-buyer scenario that may also be
interpreted to be a monopolistic price setting. One may naturally ask the
same questions in more general settings involving multiple buyers. An
immediate observation is that since our main results (Theorems \ref%
{th:infinite gap}, \ref{th:rev-m}, and \ref{th:drev}) are separations, they
apply directly also to multiple-buyer settings, simply by considering a
single \textquotedblleft significant\textquotedblright\ buyer together with
multiple \textquotedblleft negligible\textquotedblright\ (in the extreme,
with $0$-value for all goods) buyers. The issue of extending the results to
multiple-buyer settings is thus relevant to the upper bounds in the paper,
both the significant ones (Propositions \ref{th:additive-delta} and \ref%
{p:b/s<=logk}) and the simple ones (Proposition \ref{p:revm}). In this
appendix we discuss why these can all be extended to the multiple-buyer
scenario, at least if we are willing to incur a \emph{loss that is linear in
the number of buyers.} It is not completely clear where and how this loss
may be avoided.

In the case of multiple buyers, we must first choose our notion of
implementation: dominant strategy or Bayesian Nash. Also, we need to specify
whether we assume independence between buyers' valuations or allow them to
be correlated. The discussion here will be coarse enough to apply to all
these variants at the same time, with differences noted explicitly.

The next issue is how should we define the menu size in the case of multiple
buyers. In the single-buyer case we defined it as the number of options from
which the buyer may choose, which is the same as the number of allocations $%
|\{q(x):x\in \mathbb{R}_{+}^{k}\}\backslash \{(0,\ldots ,0)\}|.$ In the case
of multiple buyers, these are two separate notions. For example, consider
deterministic auctions of $k$ goods among $n$ buyers. There are a total of $%
(n+1)^{k}$ different allocations (each good may go to any buyer or to no
one), but each buyer considers only $2^{k}$ possibilities (whether \emph{he}
gets each good or not). Moreover, the set of allocations cannot be
interpreted as a menu from which the buyers may choose, since each buyer can
choose only from the possibilities offered to him (and these choices need
not be feasible overall). It takes the combined actions of all the buyers
together in order for the mechanism outcome to be determined. For this
reason we prefer to define the menu size of a multiple-buyer mechanism by
considering its menu size from the point of view of the different buyers.
Since the menu that a buyer sees is a function of the bids of the others, we
take the maximum. We thus define:

\begin{itemize}
\item An $n$-buyer mechanism has a \emph{menu size} of at most $m$ if for
every buyer $j=1,\ldots ,n$ and every $(n-1)$-tuple of (direct) bids of the
other buyers\footnote{%
Superscripts are used here for the buyers.} $x^{-j}\in (\mathbb{R}%
_{+}^{k})^{n-1}$, the number of nonzero choices that buyer $j$ faces is at
most $m$, i.e., $|\{q^{j}(x^{j},x^{-j}):x^{j}\in \mathbb{R}%
_{+}^{k}\}\backslash \{(0,\ldots ,0)\}|\leq m$.
\end{itemize}

Note that if the original mechanism was incentive compatible in dominant
strategies then the mechanism induced on player $j$ by $x^{-j}$ is also
incentive compatible. However, if the original mechanism was incentive
compatible in the Bayesian Nash sense then this need not be the case, but we
still have individual rationality\footnote{%
This assumes that the original mechanism was ex-post individually rational,
which one may verify is without loss of generality relative to ex-ante
individual rationality.} of the induced mechanism, which suffices for what
comes next.

Let us first analyze the simplest mechanisms, those with a single
non-trivial menu entry for each buyer. Clearly, bundling mechanisms satisfy
this property; however, not every mechanism that has a single non-trivial
menu entry for each buyer can be converted to a bundling mechanism. We also
need to be careful with the meaning of a bundling mechanism. Clearly, in the
case of correlated buyer valuations, the optimal mechanism for selling even
a single good (the whole bundle in our case) is not necessarily to sell it
to the highest bidder, but rather to use the bids of the others to set the
reserve price for each bidder. (Consider, for example, the case of two
buyers with a common value, where the bid of one of them should be used as
the asking price for the other.) Thus, in the rest of the discussion below
we use \textsc{BRev} to denote the optimal revenue from mechanisms that sell
the bundle only as a whole---not necessarily to the highest bidder or at a
uniform reserve price. For the case of independent buyer values, the simpler
version that sells it to the highest bidder at a fixed reserve price will
suffice as well.

What can be easily observed is that by focusing solely on the buyer that
pays the largest fraction of the revenue, we can reduce the problem to the
single-buyer case and extract at least a $1/n$ fraction of revenue by
selling the bundle to that single buyer. A full bundling mechanism can only
do better, which gives us the analog to Proposition \ref{p:revm}(i) for the
case of $n$ buyers:\footnote{%
The superscript $n$ on the various revenues denotes the number of buyers.} 
\[
\text{\textsc{BRev}}^{n}(X)\leq \text{\textsc{Rev}}_{[1]}^{n}(X)\leq n\cdot 
\text{\textsc{BRev}}^{n}(X). 
\]%
The loss of the factor of $n$ can be seen to be justified by considering
independent buyer values and the restricted definition of bundling
mechanisms already in the case of one good (i.e., $k=1$): take the
distribution where each buyer $j=1,...,n$ values the single good at $H^{j}$
with probability $H^{-j}$, and zero otherwise (independently over buyers),
for a large enough but fixed $H$.

A similar argument that focuses on the single buyer that provides the
largest fraction of revenue yields the generalization of Proposition \ref%
{p:revm}(iii) and (iv): 
\begin{eqnarray*}
\text{\textsc{Rev}}_{[m]}^{n}(X) &\leq &n\cdot m\cdot \text{\textsc{BRev}}%
^{n}(X)\text{\ \ and} \\
\text{\textsc{DRev}}^{n}(X) &\leq &n\cdot (2^{k}-1)\cdot \text{\textsc{BRev}}%
^{n}(X).
\end{eqnarray*}%
It turns out that the linear loss in $n$ is required here too, again for
independent buyer values and the restricted interpretation of bundling
mechanisms: take the construction of Theorem \ref{th:drev} for each of the $n
$ different buyers and combine it with the argument above. That is, whenever
the construction has a valuation $x$ with probability $p$, let buyer $j$
have valuation $H^{j}x$ with probability $H^{-j}p$ (independently over the
buyers).

Versions of Propositions \ref{th:additive-delta} and \ref{p:b/s<=logk} that
incur a linear loss in $n$ are also easily implied, but do not seem to be
interesting. It would seem that in both cases sharper results, in which the
additional loss due to the number of buyers is avoided, might be obtained.

\end{document}